# A magnetic monopole in pure SU(2) gauge theory

J. Smit*

Institute for Theoretical Physics, University of Amsterdam
Valckenierstraat 65, 1018 XE Amsterdam, The Netherlands

and

A.J. van der Sijs†

Theoretical Physics, University of Oxford
1 Keble Road, Oxford OX1 3NP, United Kingdom

**Abstract**

The magnetic monopole in euclidean pure SU(2) gauge theory is investigated using a background field method on the lattice. With Monte Carlo methods we study the mass of the monopole in the full quantum theory. The monopole background under the quantum fluctuations is induced by imposing fixed monopole boundary conditions on the walls of a finite lattice volume. By varying the gauge coupling it is possible to study monopoles with scales from the hadronic scale up to high energies. The results for the monopole mass are consistent with a conjecture we made previously in a realization of the dual superconductor hypothesis of confinement.

*e-mail: jsmit@phys.uva.nl

†Present address: Department of Theoretical Physics, University of Zaragoza, 50009 Zaragoza, Spain. e-mail: arjan@sol.unizar.es

# 1 Introduction

In Ref. [1] we proposed a realization of the dual superconductor hypothesis[1] [2, 3] of confinement in SU(2) gauge theory. We considered the contribution of a class of magnetic monopole configurations to the path integral, and showed that monopole condensation occurs within this class, leading to a string tension. The monopoles were of the 't Hooft-Polyakov (HP) [6, 7] form. Assuming that the string tension $\sigma$ is dominated by the contribution of these monopoles, a variational estimate yielded the value 2.3 for the fundamental ratio $\sqrt{\sigma}/\Lambda_{\overline{MS}}$. This value for the string tension was close to the results of Monte Carlo computations.

The classical magnetic monopole in SU(2) gauge theory is an analogue of the HP monopole in the SU(2) gauge-Higgs model, in particular the Bogomol'nyi-Prasad-Sommerfield (BPS) [8, 9] solution. This static classical solution is characterized by an arbitrary (mass) scale parameter $\mu$ and its mass is given by

$$M = \frac{4\pi\mu}{g^2} \,. \tag{1}$$

The role of the Higgs field in this solution is played by the timelike component $A_4$ of the vector potential in the euclidean pure gauge theory. Since this monopole corresponds to the BPS limit of zero Higgs self-coupling of the HP monopole, it is in fact a dyon.

In developing our model of confinement we made two assumptions about the properties of the monopole in the quantum theory. We proposed that quantum fluctuations induce an effective mass for the $A_4$ component of the monopole field. This would cause screening of the electric field of the dyon. Consequently, the quantum monopole was assumed to be purely magnetic at large distances. This purely magnetic nature was not expected to be crucial for the confinement itself, but it facilitated an approximate reduction of the SU(2) theory to monopole electrodynamics. The second assumption was that the mass of the monopole in the quantum theory can be written in the form

$$M = \frac{4\pi\mu}{g_R^2(\Lambda_R/\mu)}\, C(g_R^2(\Lambda_R/\mu)) \,, \tag{2}$$

with $g_R$ the running coupling in the $R$-scheme defined in terms of the quark-antiquark potential, and the function $C$ slowly increasing as for the HP monopole, $1 < C \lesssim 2$. We used $g_R$ because it naturally characterizes the interactions of electric charges $g_R/2$ and magnetic charges $4\pi/g_R$. The $\Lambda$-parameter in the $R$-scheme is related to the $\Lambda$s in the $\overline{MS}$, Pauli-Villars and conventional lattice schemes by $\Lambda_R = 1.048\,\Lambda_{\overline{MS}} = 0.964\,\Lambda_{PV} = 20.78\,\Lambda_L$ [10–13].

We then considered monopole configurations corresponding to an arbitrary scale $\mu$ in which the monopoles were separated by some minimal distance $b$. The configurations were supposed to be semiclassical, approximate, solutions of the effective action with quantum fluctations taken into account. We carried out a variational calculation which was facilitated by an approximate mapping to a compact U(1) lattice gauge theory with lattice distance $b$. The variational curve in the $b$ – $\mu^{-1}$ plane was given by $\mu b = [g_R^2(\Lambda_R/\mu)/g_R^2(b\Lambda_R)][1.59/C(g_R^2(\Lambda_R/\mu))]$. Given the assumed slow variation of $C$, the value of $\mu b$ along this curve was almost constant. With increasing $1/\mu$ along the curve, the coupling $g_R^2$ increases, the monopole mass decreases, and beyond a critical scale monopole condensation occurs, leading to confinement. The string tension was maximized leading to $\sqrt{\sigma} = 0.91 b^{-1}$ at a value of $g_R^2(b\Lambda_R) \approx 7.7$, giving the earlier mentioned value $\sqrt{\sigma} = 2.2\,\Lambda_R = 2.3\,\Lambda_{\overline{MS}}$. This value of the renormalized coupling could still be argued to be within the semiclassical domain, it corresponds to a one/two-loop beta function ratio of 0.77. For a typical guess $C = 1.25$ this gave $\mu b = 1.10$, so that the condensed monopoles were closely packed. The precise value of the monopole mass was not so important for the value of the string tension, $C = 1.59$ would give $\mu b = 1$. A change in $C$ is compensated by a change in the minimal distance $b$.

[1] The dual superconductor explanation of confinement has been criticised [4], especially the use of the so-called maximal abelian gauge and the abelian projection (for a review, see [5]). We intend to comment on the relevance of these criticisms for our work in a separate paper.

In the present paper we investigate the monopole numerically. We perform a kind of background-field calculation using lattice Monte-Carlo methods. The magnetic monopole background is induced by imposing appropriate boundary conditions. The computations reported here focus on a determination of the monopole mass as a function of its scale, to see if our assumption about the mass was justified. In particular, we determine the function $C$, defined by Eq. (2). Apart from the relevance for our model of confinement, this investigation is also interesting in its own right. We hope to address the issue of electric screening in the future.

The paper is organized as follows. In Section 2 we recall the classical magnetic monopole in the continuum. We discuss monopole boundary conditions on a space-time cylinder and the role of gauge invariance in restricting the choice of possible monopole backgrounds. In Section 3 the monopole is put on a hypercubic lattice. We verify that local minima of the lattice action correspond to continuum monopoles with small discretization errors. Section 4 deals with the monopole in the quantum theory. We describe our Monte Carlo method for calculating the monopole mass and present the results of the simulations. Section 5 contains our conclusions.

# 2 The classical magnetic monopole

The 't Hooft-Polyakov monopole [6, 7] is a solution to the equations of motion of SU(2) gauge theory coupled to a triplet Higgs field in Minkowski space. We are interested in the analogue of this monopole in pure SU(2) gauge theory in euclidean space-time.

We first recall the magnetic monopole and then discuss field configurations in a finite volume with monopole boundary conditions.

## 2.1 The monopole solutions

The Lagrangian of SU(2) gauge theory with a triplet Higgs field is given by

$$g^2\, L = -\frac{1}{4}(F^a_{\mu\nu})^2 + \frac{1}{2}(D_\mu \varphi^a)^2 - \frac{\lambda}{4}\,(\varphi^a\varphi^a - \mu^2)^2. \qquad (3)$$

The quartic Higgs potential breaks the SU(2) symmetry spontaneously to U(1) and sets the (mass) scale $\mu$ of the classical solutions. A finite energy solution has to approach

$$|\varphi| \to \mu\,, \qquad D_\mu \varphi^a \to 0\,, \qquad F^a_{\mu\nu} \to 0\,, \qquad (4)$$

fast enough at large distances. At the classical level, $\lambda$ is related to the masses of the Higgs particle and the massive vector boson by

$$2\lambda = \frac{m_H^2}{m_W^2}, \quad m_W = \mu \,. \tag{5}$$

The 't Hooft-Polyakov (HP) monopole [6, 7] is a static solution to the field equations given by the Ansatz[2]

$$\begin{aligned} A_k^a(\vec{x},t) &= \epsilon_{akl}\hat{x}_l \frac{1-K(\mu r)}{r}, & (6)\\ \varphi^a(\vec{x},t) &= \delta_{ak}\hat{x}_k \frac{H(\mu r)}{r}, & (7)\end{aligned}$$

in the "radial" or "hedgehog" gauge $A_0^a = 0$, $\vec{\varphi}(\vec{x}) \propto \vec{x}$. The monopole has magnetic charge $-1$ with respect to the residual U(1) group of gauge rotations around $\varphi^a$. In terms of the Ansatz (6,7), the field equations reduce to differential equations for the functions $H$ and $K$ with parameter $\lambda$. An analytic solution to these equations has been found only for $\lambda = 0$. It is known as the Bogomol'nyi-Prasad-Sommerfield (BPS) solution [8, 9] and given by

$$\begin{aligned} H(\mu r) &= \mu r \frac{\cosh \mu r}{\sinh \mu r} - 1 \,, & (8)\\ K(\mu r) &= \frac{\mu r}{\sinh \mu r} \,. & (9)\end{aligned}$$

For $\lambda \neq 0$, solutions have been determined numerically. Fig. 1 shows $H(x)/x$ and $K(x)$ as a function of the dimensionless parameter $x = \mu r$, for some values of $\lambda$. For $\lambda \to \infty$ the function $H(x)/x$ approaches a step function while $K(x)$ takes a limiting form close to the $\lambda = 12.5$ curve shown. A core size of the monopole may be deduced from the shape of $K$ which determines the deviation of the magnetic energy density from $1/2r^4$. In the BPS limit $\lambda = 0$ the core size may be defined as $\mu^{-1}$, while for $\lambda \to \infty$ it is smaller. Perhaps the inverse mass in units of $4\pi/g^2$, i.e. $(\mu C)^{-1}$, is a good measure for the core size. The mass (energy) of the monopole is given by

$$M = E = \frac{4\pi\mu}{g^2} C(2\lambda), \tag{10}$$

where $C(2\lambda) = C(m_H^2/m_W^2)$ is a slowly increasing function, $C(0) = 1 \leq C(2\lambda) \leq 1.787 = C(\infty)$ [14–16].

For $\lambda = 0$ the Higgs field is massless. This is reflected in the asymptotic behaviour of the Higgs field [16, 17] (the function $K(\mu r)$ decays $\propto \exp(-\mu r)$)

$$\frac{H(\mu r)}{r} \sim \mu - \frac{1}{r} - \mathcal{O}(\exp(-\mu r)) \qquad (\lambda = 0)\,, \tag{11}$$

$$\begin{aligned} \frac{H(\mu r)}{r} &\sim \mu - \mathcal{O}(\exp[-m_H r]) \qquad (0 < \lambda < 2)\\ &\sim \mu - \mathcal{O}(\exp[-\mu r]) \qquad (2 < \lambda < \infty)\,. \end{aligned} \tag{12}$$

[2]Here $r = |\vec{x}|$, $a = 1, 2, 3$ is a group index and $k, l = 1, 2, 3$ are space indices.

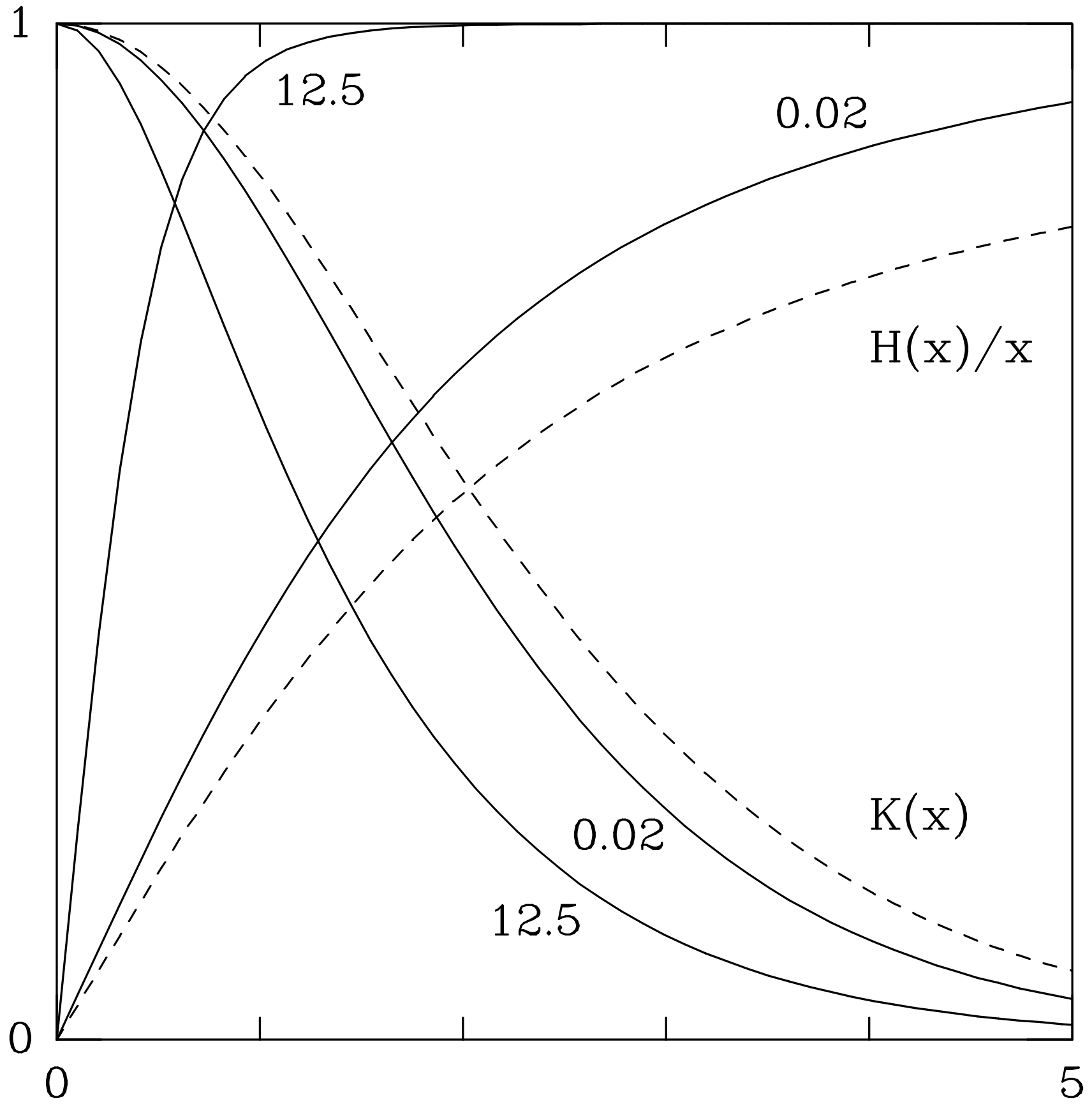

Figure 1: $H(x)/x$ and $K(x)$ for $\lambda = 0.02$ $(m_H/m_W = 0.2)$ and $\lambda = 12.5$ $(m_H/m_W = 5.0)$. The curves have been obtained numerically. The dashed curves show the $\lambda = 0$ BPS solution (8,9).

If the Higgs mass is zero there is an attractive long range interaction between a pair of (anti)monopoles, equal in magnitude to the magnetic interaction [18]. In the massive Higgs case, the Higgs interaction decays exponentially so that only the magnetic interaction survives at large distances.

We now turn to euclidean pure SU(2) gauge theory. The static HP monopole of the SU(2) gauge-Higgs model can be transferred to this theory by making the substitution $A_k^a \to A_k^a$, $\varphi^a \to A_4^a$, and the resulting configuration is a solution of the field equations,

$$\begin{aligned} A_k^a(\vec{x},t) &= \epsilon_{akl}\hat{x}_l \frac{1-K(\mu r)}{r}, \\ A_4^a(\vec{x},t) &= \delta_{ak}\hat{x}_k \frac{H(\mu r)}{r} \, . \end{aligned} \tag{13}$$

Since there is no Higgs potential the classical solution is given by the BPS solution (8,9). For the moment we will consider $H$ and $K$ as arbitrary functions, however.

The configuration (13) is a special case of the spherically symmetric Ansatz for SU(2)

gauge theory in euclidean space[3] [19],

$$
\begin{aligned}
A_k^a &= \frac{1-\phi_2}{r}\epsilon_{akl}\hat{x}_l + \frac{\phi_1}{r}(\delta_{ak} - \hat{x}_a\hat{x}_k) + A_1\hat{x}_a\hat{x}_k\,, \\
A_4^a &= A_0\hat{x}_a\,,
\end{aligned} \tag{14}
$$

with $\phi_i = \phi_i(t,r)$, $A_\alpha = A_\alpha(t,r)$. It will sometimes be convenient to consider the monopole from this perspective. In terms of this Ansatz the action becomes

$$
\begin{aligned}
S &= \int dt \int d^3x\, \frac{1}{4g^2}(F_{\mu\nu}^a)^2 \\
&= \frac{8\pi}{g^2}\int_{-\infty}^{\infty} dt \int_0^\infty dr\, [\frac{1}{2}(D_\alpha\phi_i)^2 + \frac{1}{8}r^2F_{\alpha\beta}^2 + \frac{1}{4}\frac{(1-\phi_1^2-\phi_2^2)^2}{r^2}]\,,
\end{aligned} \tag{15}
$$

where $F_{\alpha\beta} = \partial_\alpha A_\beta - \partial_\beta A_\alpha$ and $D_\alpha\phi_i = \partial_\alpha\phi_i - \epsilon_{ij}A_\alpha\phi_j$. The four-dimensional Yang-Mills theory has been transformed into a two-dimensional abelian Higgs model in a curved space with metric $g_{\alpha\beta} = r^2\delta_{\alpha\beta}$. The abelian gauge symmetry in this model comes from the U(1) group to which the original SU(2) symmetry is broken by the Ansatz (14): the abelian gauge transformation $\exp[i\omega(r,t)]$ corresponds to the $4d$ SU(2) gauge transformation

$$
\Omega(t,\vec{x}) = \exp[i\omega(t,r)\hat{x}\cdot\vec{\sigma}/2]\,. \tag{16}
$$

For four-dimensional gauge fields of the form (14) we define the "electromagnetic" $E$ and $B$ fields by the U(1) components of the field strength, given by the radial direction in SU(2),

$$
\begin{aligned}
E_i^{\rm em} &= F_{4i}^{\rm em} = \hat{x}_aF_{4i}^a = (\partial_0A_1 - \partial_1A_0)\hat{x}_i = F_{01}\hat{x}_i \\
B_i^{\rm em} &= \frac{1}{2}\epsilon_{ijk}F_{jk}^{\rm em} = \frac{1}{2}\epsilon_{ijk}\hat{x}_aF_{jk}^a = -\frac{1-\phi_1^2-\phi_2^2}{r^2}\hat{x}_i\,.
\end{aligned} \tag{17}
$$

It is clear that configurations for which $|\phi| \to 0$ at spatial infinity, such as the BPS monopole (8,9) with $\phi_1 = 0$, $\phi_2 = K$, $A_0 = H/r$, $A_1 = 0$, carry negative magnetic charge.

The BPS solution is actually self-dual. Indeed, the self-duality equations are a direct translation of the Bogomol'nyi equations of the HP monopole in the gauge-Higgs model. The exponentials in $H$ and $K$ decay over a distance of the order $\mu^{-1}$, and for $r \gg \mu^{-1}$ the electromagnetic fields (17) fall off radially as $1/r^2$. Hence, the monopole carries magnetic and electric charge and is in fact a dyon. Nevertheless, we will keep referring to it as "magnetic monopole", in view of the possibility that its electric charge is screened when quantum fluctuations are taken into account, as discussed in Ref. [1].

Finally, we note that the BPS solution with $H$ replaced by $-H$ is also a solution. This can be seen from the 'charge conjugation' invariance $A_\alpha \to -A_\alpha$, $\phi_1 \to -\phi_1$, $\phi_2 \to \phi_2$ of the action (15). This new solution has equal magnetic but opposite electric charge compared with the monopole, cf. Eq. (17). We shall call it the "anti-electric monopole".

[3]The sign of $A_\mu^a$ and $F_{\mu\nu}^a$, as well as the two-dimensional fields $A_\alpha$, $F_{\alpha\beta}$ and $\phi_i$ is opposite to that in Ref. [19]. We will use $\mu,\nu,\ldots$ for four-dimensional space-time indices, $\alpha,\beta,\ldots$ for two-dimensional space-time indices $(r,t)$ and $i,j,\ldots$ for the components of $\phi$. In two dimensions, $\partial_0 = \partial_t, \partial_1 = \partial_r$.

## 2.2 Boundary conditions on a space-time cylinder

We will consider gauge field configurations in the hyper-cylinder consisting of the spatial sphere of radius $R$ times the periodic time interval $[0,T]$, with fixed spatial boundary conditions. In a gauge field theory, it is sufficient to fix the components of the gauge potential tangential to the boundary, the normal components can be left free [20]. The path integral is a gauge-invariant functional of the fields in the boundary.

We will need boundary conditions for monopole and vacuum configurations. The choice of boundary conditions is suggested by the values of the fields in these smooth configurations. For the vacuum the choice is obvious, the tangential $A_\mu^a$ are set to zero at the boundary. For monopole configurations we impose the boundary conditions

$$\begin{aligned} A^a_{k,\mathrm{tang}} &= \frac{1}{R}\epsilon_{akl}\hat{x}_l \,, \\ A^a_4 &= \mu_0 \hat{x}_a \,. \end{aligned} \tag{18}$$

In terms of the fields in Eq. (14), $\phi_1=\phi_2=0$ and $A_0=\mu_0$ while $A_1$ remains unspecified at the boundary. These boundary conditions are compatible with a magnetic monopole of the form (13). We assume $\mu_0 R$ large enough to neglect exponential effects. We shall take $\mu_0>0$. Note that the anti-electric monopole would require $\mu_0\to-\mu_0$ in Eq. (18).

We have introduced a subscript on the parameter $\mu_0$ in the boundary conditions here, to distinguish it from the scale $\mu$ of the monopole that may be induced. The relation between $\mu_0$ and $\mu$ depends on the asymptotic behaviour of $H$. If $H(\mu r)$ and $K(\mu r)$ are given by the classical BPS solution (8,9), $\mu$ and $\mu_0$ are related by $\mu_0=\mu-1/R$. If the $1/R$-term in the large-$R$ behaviour of $H$ were absent, as in the case of the HP monopole with massive Higgs, Eq. (12), the correspondence would be $\mu_0=\mu$.

Gauss' law implies that these boundary conditions fix the magnetic charge inside the volume, for smooth configurations. They do not fix the electric charge, however, and hence are not incompatible with electric screening.

Various symmetries play a role here. First, the boundary conditions (18) are invariant under the abelian gauge transformation (16) with constant $\omega$,

$$\Omega=\exp[i\omega\hat{x}\cdot\vec{\sigma}/2]\,. \tag{19}$$

This is true regardless of the value of $R$ because $\phi_1=\phi_2=0$ in the boundary. This symmetry affects the abelian part of the gauge in which configurations appear but has no effect on $\mu_0$ and the energy.

Next, there are gauge transformations relating different values of $\mu_0$. Consider an arbitrary smooth gauge field configuration $A_\mu$ in the hypercylinder satisfying the boundary conditions (18). The gauge transformation

$$\Omega(t,\vec{x})=\exp[\pi i n\frac{t}{T}\hat{x}\cdot\vec{\sigma}]\,, \tag{20}$$

where periodicity in time requires $n$ to be even, transforms $A_\mu$ into a new configuration $A'_\mu$ with boundary conditions of the form (18) but with $\mu_0$ replaced by $\mu'_0=\mu_0+2\pi n/T$. The configuration $A'_\mu$ is singular at the origin $\vec{x}=0$, although gauge invariant quantities such as the energy density remain smooth as $\vec{x}\to 0$. Cutting out an infinitesimal ball around the singularity would avoid the singularity without affecting the energy. Eq. (20) with odd $n$ is also a symmetry of the action. This 'twisted gauge transformation' is antiperiodic, $\Omega(t = T, \vec{x}) = -\Omega(t = 0, \vec{x})$, but periodicity of the gauge fields is preserved since they transform in the adjoint representation of SU(2). Hence, this symmetry (20) relates configurations with $\mu_0$ values differing by the discrete translations

$$\mu_0 \to \mu_0' = \mu_0 + 2\pi n/T \ , \qquad n \in \mathbb{Z} \ , \tag{21}$$

but with the same energy. We shall now discuss its implications for classical monopoles in the space-time cylinder.

First, consider the idealized situation $R \to \infty$. A BPS monopole in this infinite spatial volume has scale $\mu = \mu_0 + 1/R|_{R\to\infty} = \mu_0$ and energy and action

$$E = \frac{4\pi\mu_0}{g^2} \ , \qquad S = \frac{4\pi\mu_0 T}{g^2} \ , \tag{22}$$

see Fig. 2 (solid line). The symmetry (21) shifts $\mu_0$ to other values $\mu_0'$, with action displayed by the rising dotted lines. Note that this symmetry also allows anti-electric monopoles to show up at positive values of $\mu_0$, compatible with the boundary conditions (18). They are represented by the descending dotted lines in Fig. 2. Consequently, for each value of $\mu_0$ in Eq. (18) there are various classical monopole or anti-electric monopole configurations satisfying the boundary conditions, each with different action. From this figure one can easily read off which of these configurations has lowest action, for a particular choice of $\mu_0$.

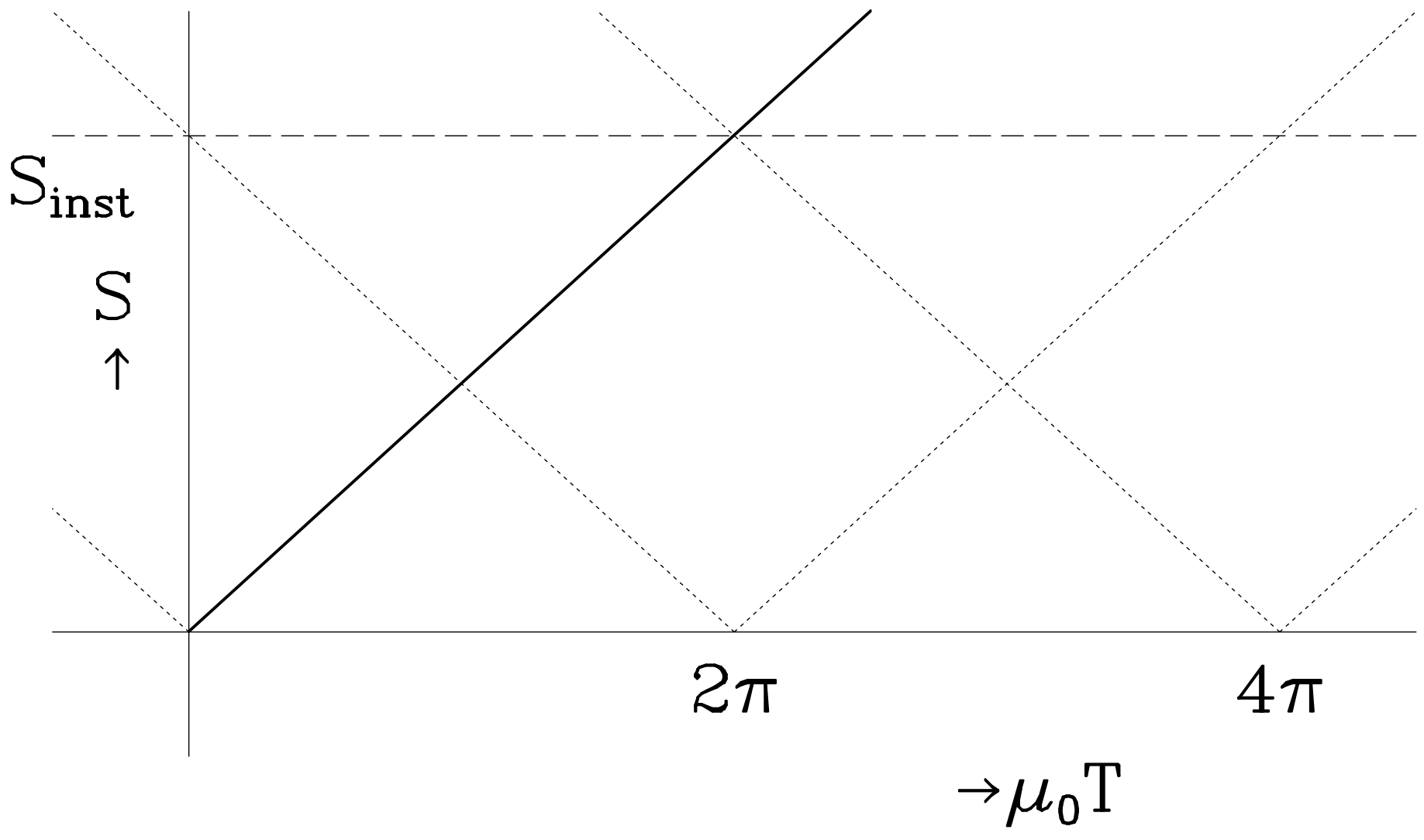

Figure 2: Action of BPS monopole configurations allowed by the boundary conditions (18) as a function of the parameter $\mu_0 T$, for a cylinder of infinite spatial radius $R = \infty$ and time extent $T$. The solid curve represents monopoles with scale $\mu_0$. Copies of monopoles and anti-electric monopoles under the symmetry (20,21) are indicated by dotted lines. The dashed horizontal line marking the instanton action, $S_{\text{inst}} = 8\pi^2/g^2$, is shown for comparison.

In the realistic case of finite cylinder radius $R$ the situation is slightly different. The boundary conditions (18) are compatible with a BPS monopole of scale $\mu = \mu_0 + 1/R$ provided $\exp(-\mu_0 R)$ is negligible, cf. Eq. (11). Such a monopole has energy $E = 4\pi\mu/g^2$, but not all of it comes from inside the cylinder. The region outside contains electric and magnetic energy adding up to

$$E_{\rm out} = \frac{1}{g^2}\int_R^\infty dr\ 4\pi r^2\,(\frac{1}{2r^4}+\frac{1}{2r^4}) = \frac{4\pi}{g^2 R}\ . \tag{23}$$

Hence the monopole energy contained inside the cylinder is

$$E_{\rm in} = \frac{4\pi\mu}{g^2} - \frac{4\pi}{g^2 R} = \frac{4\pi\mu_0}{g^2}\ . \tag{24}$$

It is amusing that the energy has this form also for a finite cylinder radius $R$, so that corrections to (24) are exponential in $\mu_0 R$.

When $\mu_0 R$ becomes small, the exponential corrections which were neglected in formulating the boundary conditions become important. The induced object is then no longer a reasonable approximation to the BPS monopole with scale $\mu = \mu_0 + 1/R$, and the energy increases relative to the value (24). Consider the limiting case $\mu_0 = 0$, for example, where the boundary conditions enforce $H(R) = 0$ and $K(R) = 0$. We determined the static minimal-action configuration inside the cylinder for these boundary conditions by numerical minimization of the energy. The solution is given by $A_0 = A_1 = \phi_1 = 0$ ($H = 0$) and $\phi_2 = K$ as shown in Fig. 3. Its energy inside the cylinder is purely magnetic and equals

$$E_{\rm in}(\mu_0 = 0) = \frac{4\pi}{g^2}\frac{1.627}{R}\ . \tag{25}$$

This solution is very well approximated by a cosine profile, for which the numerical constant in this expression is 1.631, only 0.26 % higher. As can be seen from Fig. 3, the function $K$ vanishes linearly at the boundary and the solution does not match smoothly to the outside region. For instance, the energy density jumps at the boundary. Apart from this beauty failure, this object might pass for a purely magnetic monopole with a scale of order $1/R$. Adding a magnetic outside contribution to the energy equal to half the value of Eq. (23), its total energy would be given by $E = (4\pi/g^2)(2.127/R)$.

A sketch of the monopole action as a function of $\mu_0$ in a finite cylinder is displayed in Fig. 4. The value of $S$ at $\mu_0 = 0$ and at the other minima is proportional to $T/R$, with proportionality constant given by Eq. (25). For $R \to \infty$ these minima are pushed down into cones and the limit curves are the ones shown in Fig. 2.

With Fig. 4 we can discuss the feasibility of a Monte-Carlo simulation of the magnetic monopole. For weak coupling, quantum configurations will fluctuate around the absolute minimum of the action. For a given value of the boundary parameter $\mu_0$, this minimum is readily determined from the lowest-lying solid curve in Fig. 4. Only if $0 < \mu_0 \leq \pi/T$ is this configuration a monopole of scale $\mu = \mu_0 + 1/R$, provided $\mu_0 R$ is sufficiently large that the exponential effects can be neglected. If $\mu_0 > \pi/T$, it is a monopole or anti-electric monopole with scale $\mu$ depending on the unique $\mu_0'$, following from the shift (21), for which $0 < |\mu_0'| \leq \pi/T$. It is also clear from the figure that for each classical monopole with $\mu_0$ just below $\pi/T$ (mod $2\pi/T$), there is an anti-electric monopole with parameter $2\pi/T - \mu_0$ which has almost the same energy. A simulation at such a value of $\mu_0$ may be influenced by both classical backgrounds. However, their energy densities coincide if $\mu_0 = \pi/T$.

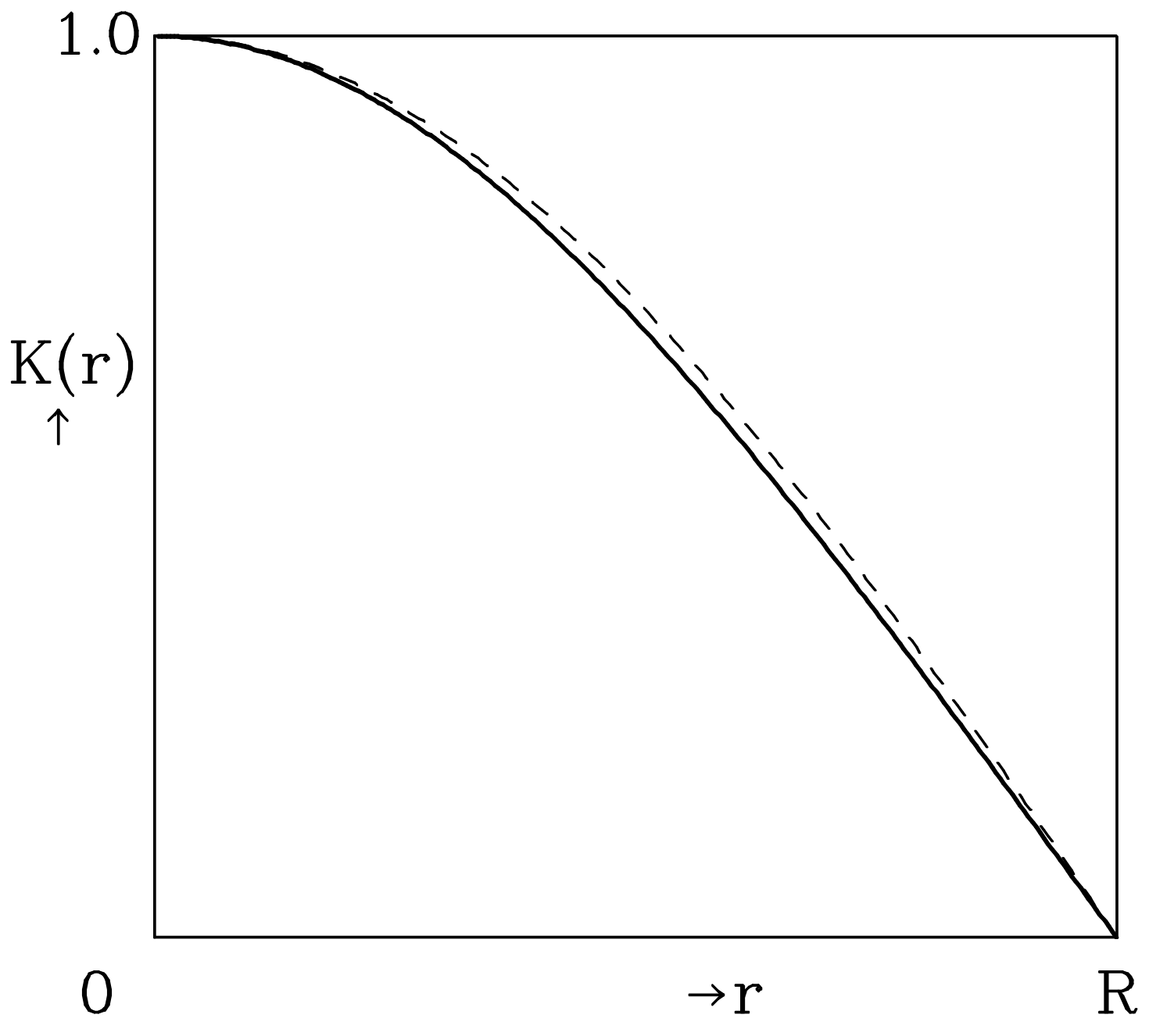

Figure 3: Solution for $K(r) = \phi_2(r)$ in the cylinder of finite radius $R$ with $\mu_0 = 0$ boundary conditions (18) (solid curve). The function $\cos(\pi r/2R)$ is shown for comparison (dashed curve).

Although it remains to be seen, in Section 3.2, how all these considerations work out on a hypercubic lattice, we draw the following tentative conclusion. The dynamical simulations will be restricted to monopole backgrounds with $0 < \mu_0 \leq \pi/T$. As we shall consider energetical aspects only, it does not matter whether we deal with a monopole or an anti-electric monopole, and there is a preference for $\mu_0 = \pi/T$. Conversely, for fixed $\mu_0$ only the temperature region $1/T \geq \mu_0/\pi$ is accessible. (Recall that in our notation $1/T$ is the temperature.) Simulations of static monopoles at zero temperature are excluded. This appears to be sufficient for application in our description of confinement, in which the monopoles were supposed to be semiclassical anyway.

# 3 The classical monopole on the lattice

In this Section we translate the considerations of Section 2 to the lattice. We describe how a monopole gauge field configuration is set up on a hypercubic lattice, discuss boundary conditions and symmetries, and check that minimization of the action leads to the classical monopole configurations.

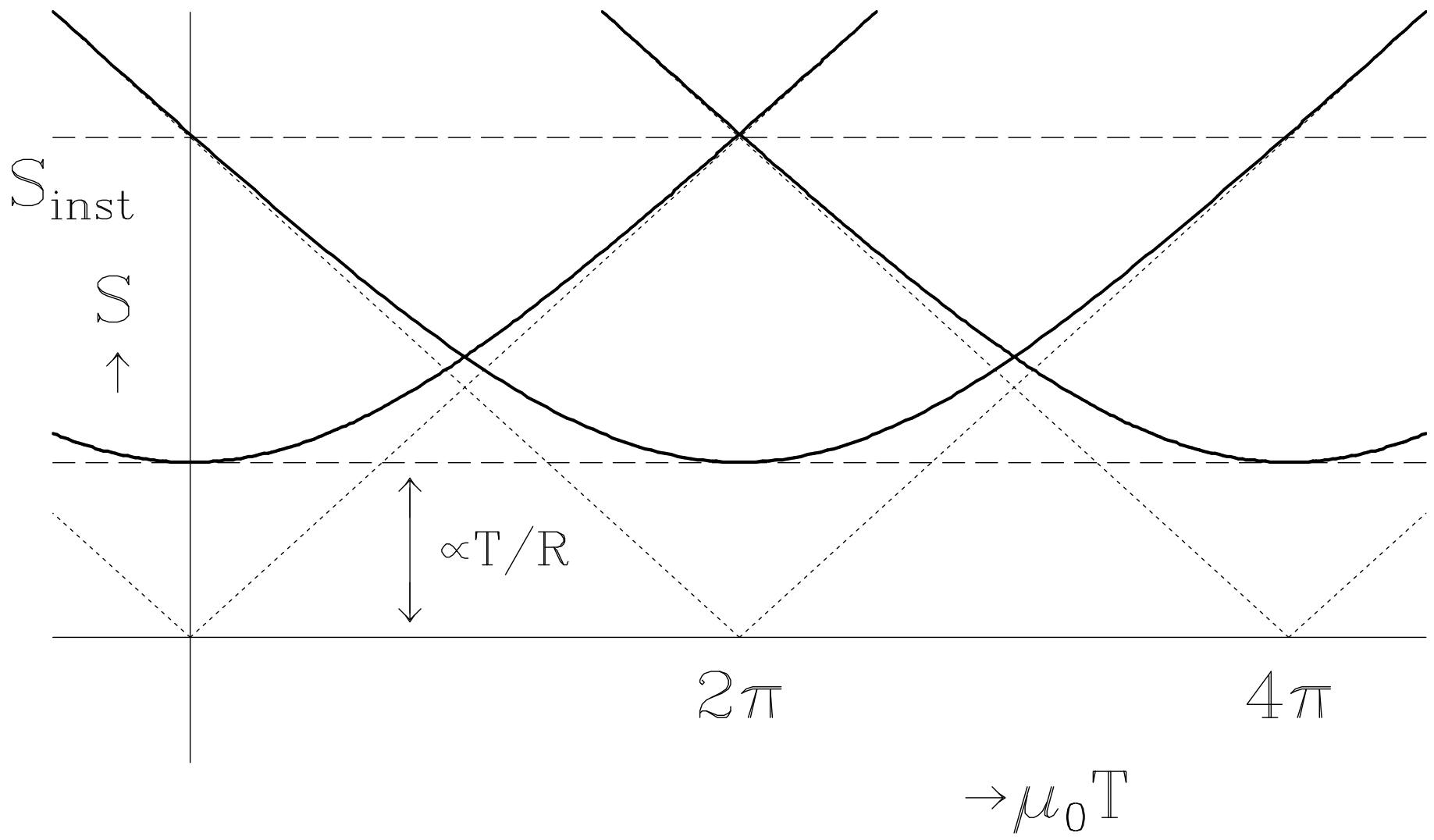

Figure 4: Inside contribution to the action of classical (anti-electric) monopole configurations and their shifted copies as a function of $\mu_0 T$, for a cylinder of finite spatial radius $R$ and time extent $T$ (solid curves). The approach to the asymptotic curves (dotted lines) is exponentially fast as $R \to \infty$.

## 3.1 The monopole on a hypercubic lattice

We use an $N^3 \times N_4$ lattice, with $N$ and $N_4$ the number of sites in the space and time directions. For the radially symmetric problem we are dealing with, a cubic lattice is not an obvious choice. However, the conventional hypercubic lattice has the advantage that it is easier to translate the Monte Carlo data into physical scales. There is a wealth of Monte Carlo data for physical quantities on hypercubic lattices to compare with.

The lattice is set up as illustrated in Fig. 5. Although there are $N$ sites in each spatial direction, the spatial volume is a cube of size $(N-1)^3 a^3$, where $a$ is the lattice spacing. We emphasize that there is no periodicity in the spatial directions and there are no links sticking out of the walls of the cubic box. We shall take $N$ to be even, so that a monopole in the middle of the box is located at the centre of an elementary lattice cube $a^3$. The time direction is periodic and has length $T = N_4 a$.

We begin with a discussion of the classical monopole on this lattice. The relation between a continuum gauge field configuration and the corresponding configuration of link fields on the lattice is given by

$$U_\mu(x) = \mathcal{P} \exp\left[-ia \int_0^1 A_\mu(x + ta\hat{\mu})\, dt\right] , \tag{26}$$

with the links in the fundamental representation,

$$A_\mu = \sum_{a=1}^{3} A_\mu^a \frac{\sigma_a}{2} . \tag{27}$$

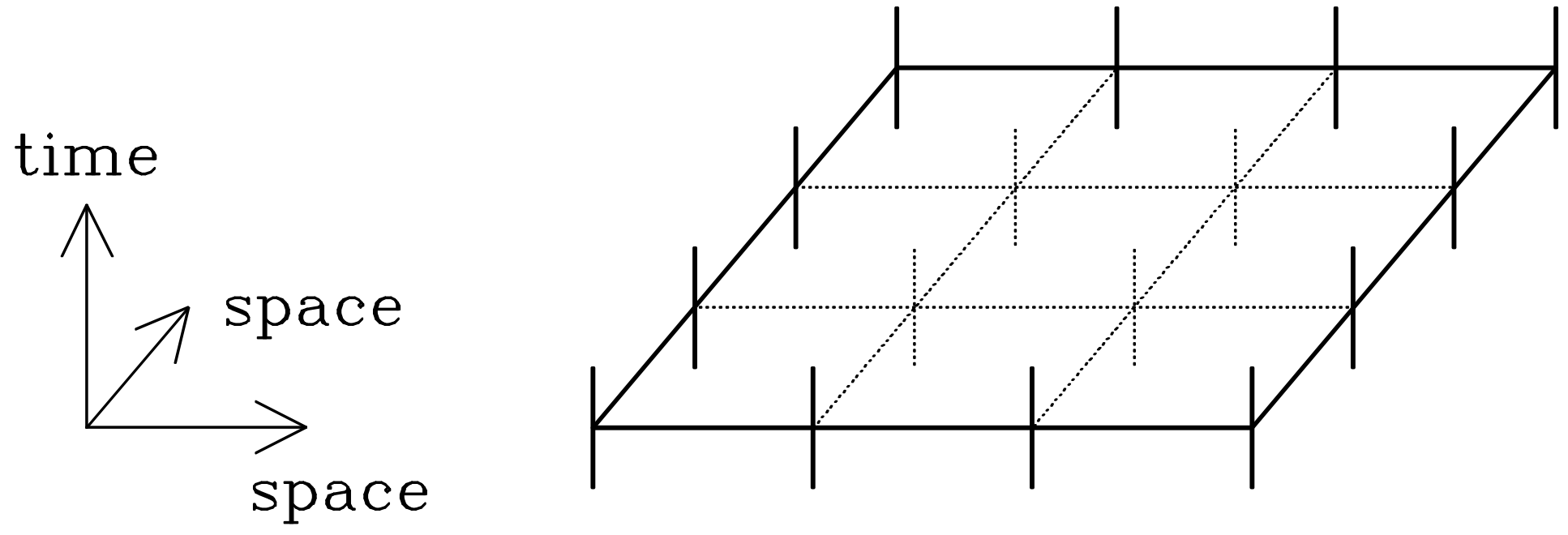

Figure 5: Illustration of the boundary conditions for an $N = 4$ lattice. Links lying in the boundary (solid lines) are fixed, dynamic links are represented by dotted lines. For clarity the timelike links have been drawn somewhat shorter.

It is important to use path-ordering (denoted by $\mathcal{P}$) to ensure that the gauge symmetry properties of the continuum gauge field are carried over to the lattice. Two continuum configurations related by a gauge transformation must lead to lattice configurations that are exactly gauge equivalent too. It is also crucial to have the correct sign in the exponent of Eq. (26). This sign is related to the sign in front of the $A \times A$ term in the nonabelian field strength $F_{\mu\nu}$ which is determined by the conventions in which the classical solution is written down. For the monopole of Eq. (13), with scale $\mu$, the link matrices (26) can be expressed in terms of conventional integrals,

$$U_k(\vec{x},t) = \exp\left[-ia\frac{\sigma_j}{2}\epsilon_{jkl}x_l\int_0^1 \frac{1-K(\mu r(s))}{r^2(s)}\,ds\right], \tag{28}$$

$$U_4(\vec{x},t) = \exp\left[-ia\frac{\sigma_j}{2}\hat{x}_j\frac{H(\mu r)}{r}\right], \tag{29}$$

$$r(s) = \sqrt{(x_k+sa)^2+\textstyle\sum_{j\neq k} x_j^2}, \qquad r = r(0).$$

In order to get an idea about discretization errors we compute the energy of the classical BPS solution (8,9), latticized in this way. In this context we think of the energy as the action per unit of time. The energy of the lattice configuration inside the box is calculated with the plaquette action

$$S = \frac{4}{g^2}\sum_{x,\ \mu<\nu}\left(1-\frac{1}{2}\mathrm{tr}\,U_{\mu\nu}(x)\right)\ . \tag{30}$$

It has to be compared with the continuum result for the energy inside the cubic box.

To get accurate results, we have to assign relative weights $< 1$ to the plaquettes in the boundary of the lattice volume, as can be seen as follows. The lattice action is defined as the sum of the plaquettes, up to a proportionality constant. However, for our finite lattice volume a more precise definition of the total action is the sum of the action contained in each of the $N_4(N-1)^3$ elementary hypercubes lying inside the volume. The action contained in one $a^4$ elementary hypercube, and assigned to the centre of this hypercube, is $\frac{1}{4}$ times the sum of the 24 plaquettes framing it. The factor $\frac{1}{4}$ comes from the fact that each plaquette contributes to 4 hypercubes. If a plaquette lies in the boundary of the lattice box, however, only 1 or 2 of its 4 hypercubes lie inside the box. Such a plaquette has to be counted with relative weight $\frac{1}{4}$ or $\frac{1}{2}$. To be specific, all the spacelike plaquettes in the boundary, i.e. of the type $U_{ij}$, get weight $\frac{1}{2}$. The timelike boundary plaquettes, $U_{i4}$, whose spacelike link lies in a face of the spatial cubic box of size $(N-1)^3a^3$ also get weight $\frac{1}{2}$, while the timelike boundary plaquettes with spacelike link lying on one of the edges of this spatial box get a weight factor of $\frac{1}{4}$. This can be verified with a counting argument: There are $3N(N-1)^2$ spacelike and $3N^2(N-1)$ timelike plaquettes per time slice in the box with the boundary included. Of the spacelike plaquettes, $6(N-1)^2$ lie in the boundary. Of the timelike plaquettes, $12(N-1)^2$ lie in the boundary, $12(N-1)(N-2)$ of them with spacelike link in a face and the remaining $12(N-1)$ with spacelike link on an edge of the $(N-1)^3a^3$ cube. Taking into account the weight factors for these boundary plaquettes, we find the expected net number of $3(N-1)^3$ spacelike and $3(N-1)^3$ timelike plaquettes contributing to the energy, at a time-slice of the box.

In the continuum, the monopole energy in a spatial box of volume $(N-1)^3a^3$ is computed by subtracting the energy contained in the region outside this box from the infinite volume value (1) for the energy of the classical monopole. For large $Na\mu$ the outside contribution can be estimated by the integral of the asymptotic energy density $1/g^2r^4$ over three-space, excluding the three-dimensional box. Denoting the length of the cubic box by $2R = (N-1)a$ and rewriting the integrand using

$$\frac{1}{r^4} = \int_0^\infty \frac{dt}{t} t^2 \exp[-r^2 t], \tag{31}$$

we get

$$\begin{aligned} E_{\text{out}} &= \frac{8}{g^2}\left(\int_0^\infty d^3\,x\,\frac{1}{r^4} - \int_0^R d^3\,x\frac{1}{r^4}\right) = \frac{2\pi^{3/2}}{g^2 R}\int_0^\infty ds[1-\text{erf}^{\,3}(s)] \\ &= \frac{4\pi}{g^2\,\alpha R}\,, \qquad \alpha = 1.2031\ . \end{aligned} \tag{32}$$

Thus we obtain for the energy inside the box

$$E_{\text{in}} = E - E_{\text{out}} = \frac{4\pi}{g^2}\left(\mu - \frac{1}{\alpha R}\right). \tag{33}$$

By comparing this result with the computed plaquette energy for various lattice sizes, at fixed $a\mu$, and extrapolating to large $\mu R$, we can estimate the discretization error in the energy. For $a\mu = \pi/8$, this discretization error is below the 1% level. Even for $a\mu$ as large as 1.0 the discretization error is only 2%.

Another check of discretization effects is provided by cooling (minimization of the action) of the latticized classical monopole, keeping the fields in the boundary fixed at their classical values, to find the corresponding local minimum of the lattice plaquette action. On an $N = 16$ lattice, for example, the energy of an $a\mu = 1.0$ monopole got lowered less than 0.1%. We conclude that the lattice version of the classical monopole is smooth enough on reasonable lattices such that discretization errors in the plaquette energy are small.

## 3.2 Boundary conditions on the lattice

As discussed in Section 2.2, boundary conditions are imposed by fixing the gauge field components tangential to the boundary, cf. Eq. (18). On our lattice, this is a very natural procedure, see Fig. 5. We fix the spacelike links lying in the boundary of the cubic spatial volume and the timelike links originating at sites in the boundary, at each time slice. The boundary conditions in the time direction are taken periodic.

The links lying in the boundary are fixed according to the appropriate link matrices (28,29), with exponential terms neglected. For the spatial links, this means $K \to 0$, as in (18). For the timelike links, we set $H/r \to \mu - 1/r$. Due to the loss of spherical symmetry, $r$ varies in the boundary, so the result is different from (18), where $\mu - 1/r = \mu_0$ was fixed. Actually, we shall set $H/r \to \mu_1 - 1/r$, in order to be able to distinguish between the boundary parameter $\mu_1$ and the induced monopole scale $\mu$. We shall also study boundary conditions with $H/r \to \mu_0$, i.e. with fixed $\mu_0$. These may lead to approximately the same results if $r$ may be considered effectively constant in the boundary, such that $\mu_0 = \mu_1 - 1/R_{\rm eff}$.

Summarizing, we shall distinguish two kinds of qualitatively different boundary conditions. We will call them BPS boundary conditions ($K = 0$, $H/r = \mu_1 - 1/r$),

$$
\begin{aligned}
A^a_{k,\,\rm tang} &= \frac{1}{r(\vec{x})}\,\epsilon_{akl}\hat{x}_l \;, \\
A^a_4 &= (\mu_1 - \frac{1}{r(\vec{x})})\,\hat{x}_a \;,
\end{aligned}
\tag{34}
$$

and HP boundary conditions ($K = 0$, $H/r = \mu_0$),

$$
\begin{aligned}
A^a_{k,\,\rm tang} &= \frac{1}{r(\vec{x})}\,\epsilon_{akl}\hat{x}_l \;, \\
A^a_4 &= \mu_0\,\hat{x}_a \;.
\end{aligned}
\tag{35}
$$

These names are inspired by the behaviour of $H/r$ for the massless Bogomol'nyi-Prasad-Sommerfield solution (8,11) and the massive-Higgs version of the 't Hooft-Polyakov monopole (12), respectively. Also in this cubic volume the tangential boundary conditions are not incompatible with electric screening.

If we adopt HP boundary conditions (35), the boundary links become

$$
U_2(\vec{x}) = \cos\frac{\chi(\vec{x})}{2} - i\,\hat{n}(\vec{x})\cdot\vec{\sigma}\sin\frac{\chi(\vec{x})}{2} \tag{36}
$$

$$
= \sqrt{\frac{1}{2}\left(1 + \frac{r^2 + ya}{r\sqrt{r^2 + 2ya + a^2}}\right)} - i\hat{n}\cdot\vec{\sigma}\sqrt{\frac{1}{2}\left(1 - \frac{r^2 + ya}{r\sqrt{r^2 + 2ya + a^2}}\right)}, \tag{37}
$$

$$
\chi(\vec{x}) = \arctan\left(\frac{a\sqrt{x^2 + z^2}}{r^2 + ya}\right), \tag{38}
$$

$$
\hat{n}(\vec{x}) = \left(\frac{z}{\sqrt{x^2 + z^2}}, 0, -\frac{x}{\sqrt{x^2 + z^2}}\right) \tag{39}
$$

(and similarly for $U_1$ and $U_3$),

$$
U_4(\vec{x}) = \cos\frac{a\mu_0}{2} - i\,\hat{x}\cdot\vec{\sigma}\sin\frac{a\mu_0}{2} \,. \tag{40}
$$

Here $\chi(\vec{x})$ is the angle between the vectors pointing from the origin to the beginning and end of the link. In the case of BPS boundary conditions (34), the spatial links are the same as in (36–39), whereas the timelike links are obtained by replacing $\mu_0 \to \mu_1 - 1/r(\vec{x})$ in (40).

For HP boundary conditions the timelike plaquettes in the boundary are exactly zero. For $U_{24}$ for example,

$$U_2^\dagger(x,y,z)\,U_4(x,y,z)\,U_2(x,y,z) = U_4(x,y+1,z)\,. \tag{41}$$

This agrees with the continuum result that the electric field is screened in the HP case, and hence zero if exponential effects are neglected. Similarly, one can show that for BPS boundary conditions the timelike plaquettes equal $1-\cos(a/2r_1 - a/2r_2)$, independent of $\mu_0$. Here $r_{1,2}$ are the values of $r$ at the sites between which the plaquette is located. Since the argument of the cosine is very small, $\mathcal{O}(a^2/R^2)$, this electric energy density is nearly equal to the continuum value $\propto 1/r^4$.

We have to assess the role of the shift symmetry (21) in our cubic lattice geometry. In the continuum it comes from the gauge transformation (20) and its twisted (antiperiodic) generalization. The normal (periodic) gauge transformation (20) can be taken over on the lattice without change. However, on the lattice we cannot use the twisted version because the link variables $U_\mu(x)$ are not invariant under the centre of the gauge group. The role of twisted gauge transformations is taken over by (normal) gauge transformations combined with the operation of multiplying each timelike link on a particular time slice by an element of the centre $(-I)$, which is a symmetry of the action and the integration measure.

The transformation (20) applied to a smooth continuum configuration introduces a singularity in the spatial origin. On our lattice, the origin is located in the centre of an elementary lattice cube, so the only remnant of this singularity is that the link fields close to the origin get far away from the unit matrix. However, on the lattice we need not restrict ourselves to smooth gauge transformations. We can in fact restrict the transformations to the boundary, using the identity transformation in the interior.

On the lattice, this gives rise to the same exact shift symmetry (21) as in the continuum, whether we adopt HP or BPS boundary conditions. This is guaranteed by the fact that $\phi_1 = \phi_2 = 0$ at the boundary, in the terminology of Eq. (14), so that $\vec{\phi}$ is invariant under the transformation, which acts on it as a rotation. (For the same reason, the abelian symmetry (19) of the boundary conditions also holds on our lattice.) The transformation of $A_0$ at the boundary is given by

$$\mu_1 - \frac{1}{r(\vec{x})} \;\to\; \mu_1 + \frac{2\pi n}{T} - \frac{1}{r(\vec{x})} \qquad \text{(BPS)}\,, \tag{42}$$

$$\mu_0 \;\to\; \mu_0 + \frac{2\pi n}{T} \qquad \text{(HP)}\,, \tag{43}$$

while $A_1$ does not enter into the boundary conditions. In both cases, the transformation induces a constant shift in $\mu_0$ or $\mu_1$, as in (21), even for the BPS case where the value of $A_0$ depends on the boundary coordinates.

Next, we consider the role of the anti-electric monopole. It is obtained from the monopole by mapping $H/r \to -H/r$, at each value of $\vec{x}$. In the case of HP boundary conditions, this means that $\mu_0$ is mapped into $-\mu_0$ in the boundary, so that we end up with HP boundary conditions for a different $\mu_0$-value. Consequently, we envisage behaviour similar to that displayed in Fig. 4, the monopole having the same energy as the corresponding anti-electric monopole. For BPS boundary conditions, the mapping to the anti-electric monopole means $\mu_1 - 1/r(\vec{x}) \to -\mu_1 + 1/r(\vec{x})$ in the boundary. Because of the sign change in the $1/r(\vec{x})$ term the anti-electric monopole does not obey the BPS boundary conditions (34). However, in sofar as BPS boundary conditions are approximated by HP boundary conditions, we expect that anti-electric monopoles can also be induced by BPS boundary conditions. This will be illustrated by the following computer simulations.

We took an $8^3 \times 12$ lattice, and looked for local minima of the action for HP or BPS boundary conditions at given values of $\mu_0$, $\mu_1$. A random ('hot') starting configuration was created and its action was iteratively minimized ('cooled') subject to the boundary conditions. The actions of the resulting configurations were plotted as a function of $\mu_0$ and $\mu_1$. This was done for five different hot starting configurations, with identical boundary conditions, to allow different local minima to appear. The procedure was carried out for HP boundary conditions with $\mu_0$ ranging from 0 to $2\pi/T$ and for BPS boundary conditions with $\mu_1$ between 0 and $4\pi/T$.

Fig. 6 shows the resulting scatter plots of the energy (i.e., action per unit of time). The five points plotted at each value of $\mu_0$, $\mu_1$ are distributed among several local minima of the energy. We would like to identify them with the classical monopole solutions of the various symmetry-related scales discussed earlier. Thereto, we shifted the $\mu_0$ and $\mu_1$ values towards the appropriate values $\mu_0'$ and $\mu_1'$, according to (42,43), and replotted the data in Fig. 7. For HP boundary conditions ($a$), where $\mu_0'$ and $-\mu_0'$ give the same energy, the points are plotted as a function of $|\mu_0'|$. The BPS data ($b$) are plotted against $\mu_1'$ itself.

A striking feature in Figs. 6$b$ and 7$b$, for BPS boundary conditions, is the apparent reflection symmetry around $\mu_1 T/2\pi \approx 0.5$ (mod 1) and $\mu_1' T/2\pi \approx 0.5$, respectively. This suggests that the configurations on the descending branches of the curves in these figures can be interpreted to a good approximation as anti-electric monopoles. A quadratic fit demonstrates that the symmetry is almost perfect, see Fig. 7$b$. The central value is $\mu_1' T/2\pi = 0.481$, which means $a\mu_1' = 0.252$. We note that Figs. 7$a$ and $b$ almost coincide when $a\mu_1'$ is shifted over 0.252, as is illustrated in Fig. 8. This confirms our expectation that, as far as the energy is concerned, HP boundary conditions are to a good approximation equivalent to BPS boundary conditions, with

$$\mu_0 = \mu_1 - 1/R_{\text{eff}} \quad (\text{mod } 2\pi n/T), \tag{44}$$

where $a/R_{\text{eff}} = 0.252$. Since $R_{\text{eff}} \propto R$, because it appears to be independent of $\mu_{0,1}'$, it follows from $R/a = (N-1)/2 = 3.5$ that

$$R_{\text{eff}} = 1.13\, R. \tag{45}$$

Figs. 6$a$ and 7$a$, for HP boundary conditions, confirm the picture sketched in Section 2.2, cf. Fig. 4. Because of the HP boundary conditions together with the cubic geometry, we expect $1/R$ corrections to the asymptotic behaviour $E_{\text{in}} \propto \mu_0$, instead of only exponential corrections as in the cylindric case, cf. Eq. (24). In fact (omitting the primes on the $\mu$'s now),

$$E_{\text{in}}^{\text{HP}}(\mu_0) \approx E_{\text{in}}^{\text{BPS}}(\mu_1) = \frac{4\pi}{g^2}(\mu_1 - \frac{1}{\alpha R}) = \frac{4\pi}{g^2}\left(\mu_0 + \frac{1}{R_{\text{eff}}} - \frac{1}{\alpha R}\right) = \frac{4\pi}{g^2}\left(\mu_0 + \frac{0.051}{R}\right), \tag{46}$$

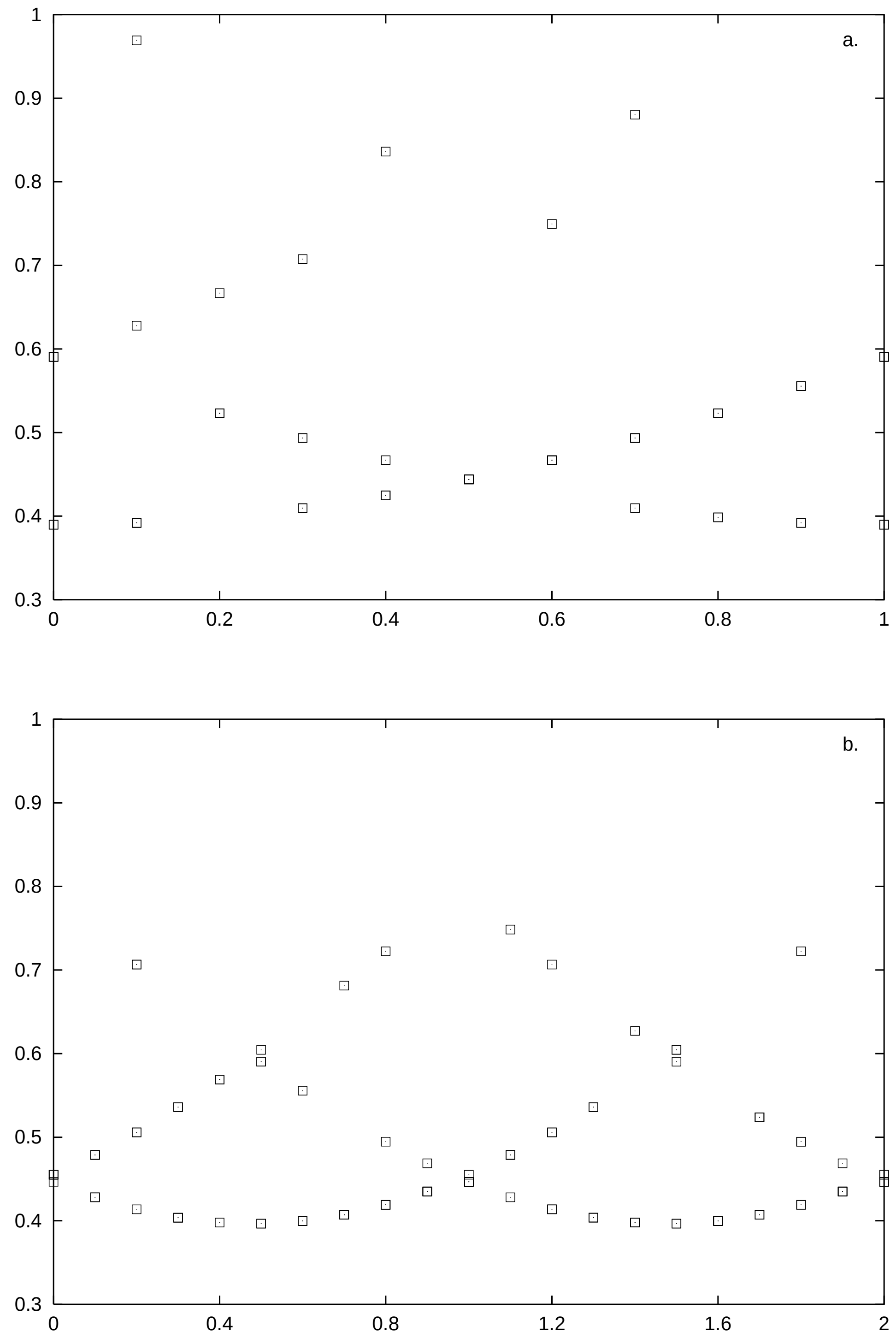

Figure 6: Energy, in units of $4\pi/ag^2$, of local minima found on the $8^3 \times 12$ lattice as a function of $\mu_0 T/2\pi$, where $T = 12$, for HP boundary conditions $(a)$, and as a function of $\mu_1 T/2\pi$ for BPS boundary conditions $(b)$. For each value of $\mu_0$, $\mu_1$, five points are plotted, several of which coincide.

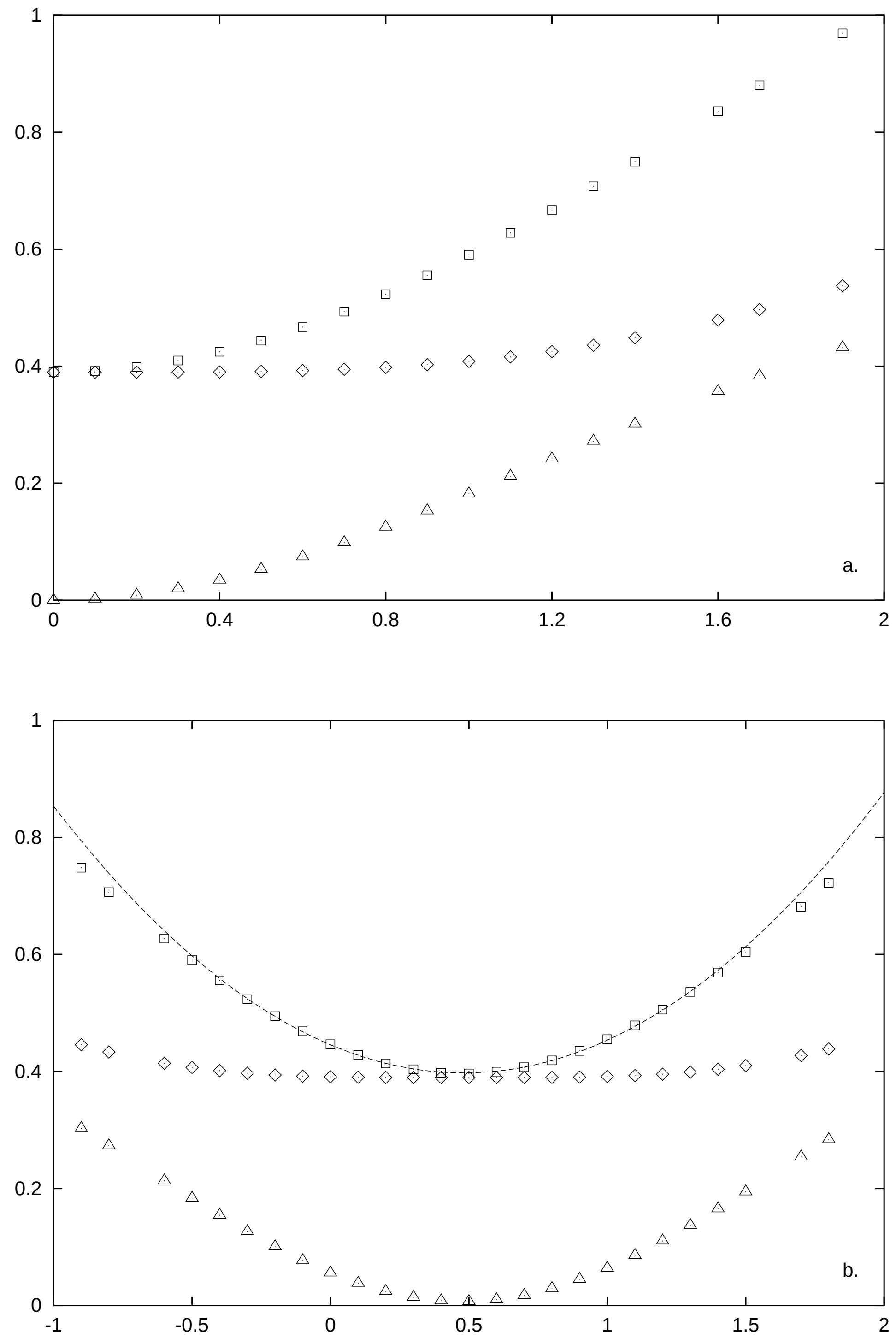

Figure 7: The energies of Fig. 6, now plotted as a function of $\mu_0' T/2\pi$ (a), or $\mu_1' T/2\pi$ (b), where $\mu_0'$, $\mu_1'$ are obtained from $\mu_0$, $\mu_1$ through the appropriate shift (42,43). The HP case is depicted in *a*, BPS in *b*. For HP boundary conditions (*a*), the monopole (with $\mu_0'$) and the anti-electric monopole (with $-\mu_0'$) have both been plotted at positive $\mu_0'$, confirming that their energies are equal. In addition to the total energy ($\Box$), the magnetic ($\Diamond$) and electric ($\triangle$) contributions are shown separately. The dashed line in *b* is a quadratic fit.

and the $\mathcal{O}(1/R)$ correction to the energy formula (24) for the cylindrical case is small. It is also satisfying that the minimum $(g^2/4\pi)\, aE_{\rm in} \approx 0.40$ in Figs. 6-8 can be understood from (25) in terms of $R_{\rm eff}$: $1.627\, a/R_{\rm eff} = 0.41$.

We conclude that, with the appropriate modifications and the introduction of $R_{\rm eff}$, the analysis in Sect. 2 can be taken over to the cubic lattice.

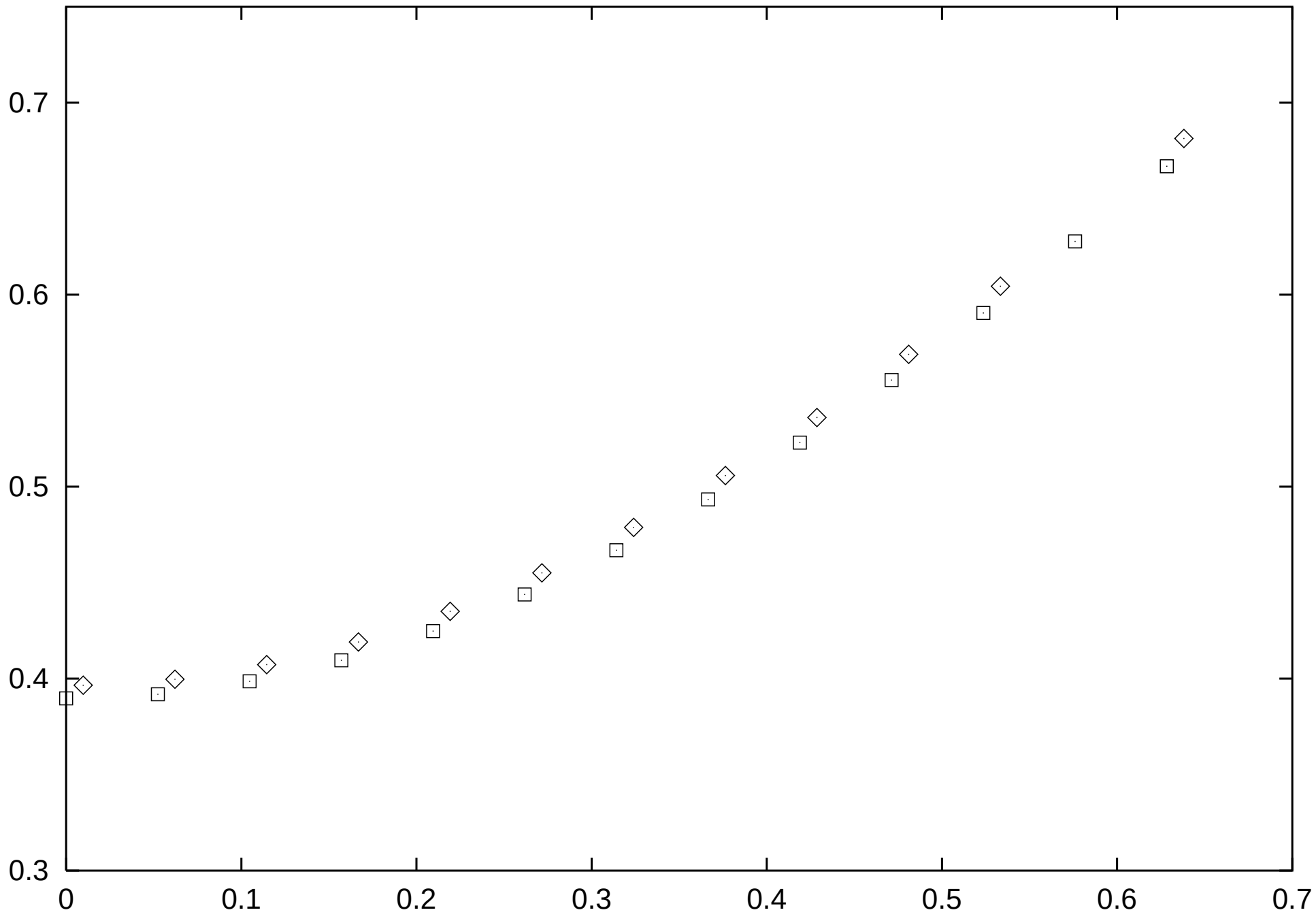

Figure 8: The data of Figs. 7 $a$ and $b$, combined into one graph (total energy only), now as a function of $a\mu_0'$ (a few points have been left out). The BPS data ($\Box$) have been shifted over a distance $1/R_{\rm eff}$ along the horizontal axis. The HP points are plotted as diamonds ($\Diamond$) here.

# 4 Monte Carlo simulations for the quantum monopole

The quantum theory is described by the euclidean path integral

$$Z[\beta, \{U_{\rm bnd}\}] = \int DU_{\rm int}\, \exp[-S(\beta, \{U_{\rm bnd}\}, \{U_{\rm int}\})]\ . \tag{47}$$

Here $\{U_{\rm bnd}\}$ and $\{U_{\rm int}\}$ are the sets of fixed boundary links and dynamical interior links, respectively. The integration symbol stands for a group invariant integration for each interior link $U_{\rm int}$. This path integral is a gauge-invariant functional of the link fields in the boundary. This is reminiscent of the gauge-invariant background field dependence of the path integral and the effective action in the background field method.

We have seen that the monopole boundary conditions induce classical monopoles as local minima of the action in a $(2R)^3 \times T$ lattice hypercube. Recall that $R = a(N-1)/2$, $T = aN_4$ on our $N^3 \times N_4$ lattice. The appropriate boundary conditions for the pure gauge theory are BPS, given in (34). It turned out that these are equivalent, to a good approximation, to HP boundary conditions (35). We may therefore restrict ourselves to HP boundary conditions in the dynamical simulations to follow. This is fortunate since they would be more appropriate, if our assumption is correct that the classical dyon turns into a purely magnetic monopole in the quantum theory.

As already mentioned in Sect. 2, for the Monte Carlo computations of the quantum monopole the global minimum of the action is important. For the $8^3 \times 12$ example in Fig. 6 this corresponds to the envelope of the lowest data points. It follows that we may restrict the values of $\mu_0$ to $0 \leq \mu_0 \leq \pi/T$.

Figs. 6,7 show rather large finite box effects for these values of $\mu_0$. Classically we can eliminate such effects by using the exact BPS values for the boundary conditions (including 'exponential effects'), as in Sect. 2.1, but in the quantum theory we do not know the precise form of the monopole. This is why we use the universal long distance form of the monopole fields in formulating the BPS and HP boundary conditions. To reduce the finite box effects we shall use the largest value $\mu_0 = \pi/T$ in the numerical simulations. Note that even then $\mu_0 R$ is only 0.92 for the $8^3 \times 12$ lattice in Fig. 6.

Having chosen HP boundary conditions for the Monte Carlo simulations, characterized by the boundary parameter $\mu_0$, we note that the $\mu$-scale of the monopole depends on the nature of the quantum monopole. If it stays BPS-like, then $\mu = \mu_0 + 1/R_{\rm eff}$, as in the classical theory. If, on the other hand, it becomes HP-like in the quantum theory, then its $\mu$-scale is $\mu = \mu_0$. So we have

$$a\mu = a\mu_0 = \pi/N_4, \quad \mu R = (\pi/2)(N-1)/N_4 \qquad \text{(HP)}, \quad (48)$$
$$a\mu = a\mu_0 + 1/(1.13 * (N-1)/2), \quad \mu R = (\pi/2)(N-1)/N_4 + 1/1.13 \quad \text{(BPS)}, \quad (49)$$

according to the quantum monopole turning out HP-like or BPS-like. In the classical regime, at weak coupling, we know the monopole behaves BPS-like. At stronger coupling, we will have to keep both options open for our analysis of the results of the simulations.

The finite box effects can be decreased by increasing $\mu_0 R$, i.e. increasing the ratio $N/N_4$. However, we will have to deal with $\mu_0 R$ values that are not very large, for the following reasons. Firstly, in the quantum case we should not take $R$ too large if we perform nonperturbative computations. The reason is that nonperturbative effects generate a finite correlation length, as given by the glueball mass or the string tension, implying that boundary conditions become irrelevant for large volumes. To stay in the semiclassical regime we have to use a finite volume, with $R\sqrt{\sigma}$ smaller than some number $z$ of order 1. In the monopole description of confinement this restriction can be interpreted by saying that we would like to study one monopole, which may include the quantum fluctuations of other small scale objects, such as instantons or monopoles, but which should exclude nonperturbative effects of other large scale monopoles, as these destroy the monopole under study. Secondly, we want to study monopoles with inverse scales up to $\mu^{-1}\sqrt{\sigma} \approx 1$. Since $\mu^{-1}\sqrt{\sigma} \leq \mu_0^{-1}\sqrt{\sigma} \leq z/\mu_0 R$, we have a conflict between large $\mu_0 R$ and getting $\mu^{-1}\sqrt{\sigma} \approx 1$.

We recall that the value of the lattice distance $a$, e.g. in units of the string tension $\sqrt{\sigma}$, $a\sqrt{\sigma}$, is controlled by the bare coupling constant $\beta = 4/g^2$. On a given $N^3 \times N_4$ lattice, decreasing $\beta$ from infinity means increasing $R\sqrt{\sigma}$ and $\mu^{-1}\sqrt{\sigma}$ from zero onwards, keeping $\mu R$ constant. In this way we can access sizes up to $R\sqrt{\sigma}$ of order 1, while it depends on the value of $\mu R$ if we can also reach $\mu^{-1}\sqrt{\sigma} \approx 1$.

Here we shall use a symmetric lattice, $N = N_4$, as a compromise between getting sufficiently large $\mu^{-1}\sqrt{\sigma}$ and not too small $\mu R$. The resulting values of $\mu R$ are rather small, cf. (48,49). Consequently our Monte Carlo simulations deal mainly with the interior of the monopole. For HP, $\mu R$ is even smaller than for BPS. On the other hand HP-like monopoles are expected to be smaller, at given $\mu$, than BPS-like monopoles, so their interior may still fit reasonably well in the box.

## 4.1 Strategy for calculating the monopole mass

We write the monopole mass as

$$M = M_{\text{in}} + M_{\text{out}} \,, \tag{50}$$

where $M_{\text{in}}$ is the contribution to the mass which will be computed in the Monte Carlo simulation and $M_{\text{out}}$ comes from the infinite region outside. As discussed above, the latter cannot be dealt with nonperturbatively, it makes sense only in an analytic semiclassical approximation. Here we shall take it into account as follows.

In the pure SU(2) gauge theory the classical monopole is BPS, but in the quantum case we shall explore the possibility that it turns HP-like and analyse our results both for BPS and for HP behavior. For a classical BPS monopole the electric and magnetic fields decay as $1/r^2$ at large distances. For an HP monopole the magnetic field decays as $1/r^2$ while the electric field is exponentially small. In the box the outside contribution to the energy is $1/\alpha R$ (BPS) or $1/2\alpha R$ (HP), in units of $4\pi/g^2$, where $\alpha = 1.2031$ is the correction factor introduced in Eq. (32). In the quantum theory the outside magnetic field is expected to remain that of a magnetic monopole of charge $-1$, whereas the expected electric field corresponds to electric charge 1 in the BPS case, and to charge 0 in the HP case.

Thus we write for the outside contribution to the monopole mass

$$M_{\text{out}}(\mu) = \frac{4\pi\mu}{g_R^2(\Lambda_R/\mu)} C_{\text{out}}(\mu R), \quad C_{\text{out}} = \frac{1}{\alpha\mu R} \quad \text{(BPS)} \,, \tag{51}$$

$$= \frac{1}{2\alpha\mu R} \quad \text{(HP)} \,. \tag{52}$$

Presumably it would not matter very much if we use the renormalized coupling in other reasonable schemes such as $\overline{MS}$ or Pauli-Villars, because the corresponding $\Lambda$ scale parameters are very close, and furthermore (51,52) is supposed to be a correction to the total mass. On the other hand, our calculation of $M_{\text{in}}$ is scheme independent, and a change in the definition of the coupling is compensated by a change in $C_{\text{in}}$.

We define the monopole mass $M_{\text{in}}$ for the interior region by

$$M_{\text{in}} = -\frac{1}{T} \ln \frac{Z_{\text{mon}}(\beta; a\mu_0)}{Z_{\text{vac}}(\beta)} \,. \tag{53}$$

Here $Z_{\text{mon}}(\beta; a\mu_0)$ is the partition function (47) for a monopole background subject to boundary conditions specified by the parameter $\mu_0$ and $Z_{\text{vac}}(\beta)$ is the analogous definition for vacuum boundary conditions, i.e. all $A_\mu^a = 0$. Eq. (53) defines the monopole mass as the free energy of the monopole, with the vacuum free energy subtracted. In this way, the exponentiated monopole mass expresses the effective Boltzmann weight of the monopole in the path integral. In our simulations we will also monitor the internal energy $\Delta E$,

$$\Delta E = \frac{1}{T}(\langle S\rangle_{\rm mon} - \langle S\rangle_{\rm vac}), \tag{54}$$

and its density as a function of $r = |\vec{x}|$.

We assume the mass (53) to be ultraviolet finite. Renormalization properties of functional integrals depending on boundary conditions are studied in [21], but the results there are not directly applicable to our case. Reassuring to us are also the following observations. The $A_\mu$ are finite in perturbation theory in the sense that the divergent parts of the wavefunction and coupling constant renormalizations cancel. (We use the convention in which $A_\mu$ 'contains' the bare coupling constant $g$. We have of course employed the same convention for the boundary values, e.g. $\mu_0$ does not require renormalization). Furthermore the path integral $Z_{\rm mon}$ depends continuously on the fields in the boundary. These boundary fields can be sent to zero in a continuous fashion, leading $Z_{\rm mon}$ towards $Z_{\rm vac}$ and the monopole mass to zero. (Also in the continuum the boundary fields can be deformed to zero continuously.) Note that in the limit $\beta \to \infty$ the monopole mass assumes its classical value.

The mass (53) can be expressed in terms of expectation values of observables, making it accessible to Monte Carlo computation. Differentiating Eq. (53) we find (recall $S \propto \beta \equiv 4/g^2$)

$$\frac{\partial M_{\rm in}(\beta)}{\partial\beta} = \frac{1}{\beta}\Delta E(\beta)\,. \tag{55}$$

This leads to

$$M_{\rm in}(\beta) = \int_0^\beta \frac{d\tilde\beta}{\tilde\beta}\Delta E(\tilde\beta)\;, \tag{56}$$

where the integration constant corresponds to $M_{\rm in} = 0$ at $\beta = 0$. This is clear from $S \propto \beta$ and the normalized measure for the integration over the interior links.

We use the trapezoid rule in computing

$$aM_{\rm in}(\beta) = a\sum_{i=1}^{N}\frac{1}{2}(\beta_i - \beta_{i-1})(\Delta E(\beta_i) + \Delta E(\beta_{i-1}))\;, \tag{57}$$

for a set of values $\beta_0 = 0 < \beta_1 < \beta_2 \ldots < \beta_{N-1} < \beta_N = \beta$ such that (57) is a good approximation to (56). This requires two Monte Carlo simulations at each value of $\beta_i$, one for monopole and one for vacuum boundary conditions.

## 4.2 The internal energy

We shall present the results of a simulation on an $8^4$ lattice. Recall that this means our spatial box has volume $(2R)^3 = (7a)^3$ and the length of the periodic time direction is $T = 8a$. For convenience we collect here the numerical values of some of the quantities introduced earlier. We have $R_{\rm eff}/a = 1.13\,R/a = 3.96$. For HP, $a\mu = \pi/8 = 0.39$ and $\mu R = 1.37$. For BPS, $a\mu = \pi/8 + a/R_{\rm eff} = 0.65$ and $\mu R = 2.26$.

The simulations were carried out as follows. We fixed the boundary conditions appropriately and took a hot (random) start in the interior of the lattice. We started at the lowest $\beta$-value, then stepwise increased $\beta$ to $\beta_2$, $\beta_3$, etc. up to $\beta_N$, and subsequently we decreased $\beta$ again, down to $\beta_1$. This allowed us to check for possible "hysteresis" effects, due to unsufficient thermalisation. As a matter of fact, we carried out two such cycles. The second one was done to obtain data at intermediate $\beta$-values where more accuracy appeared necessary after the first run. See the specifications detailed in Table 1. At each $\beta$-value we started with 400 thermalisation sweeps and subsequently 4000 measurements of the action were done, separated by 2 sweeps. The statistical errors on $\Delta E$ were determined by pointwise subtracting the 2 sets of 4000 numbers (using the same pseudo-random numbers for monopole and vacuum boundary conditions) and binning the 4000 differences. The errors on $aM_{\text{in}}$ were determined similarly.

Our SU(2) program used the site labelling and "gather and scatter" method described in Ref. [22], and additional logical vectors, used as masking vectors throughout the program, controlled the special behaviour of the links in the boundary, see Fig. 5. The pseudo-heatbath method [23, 24] combined with $\omega = 2$ overrelaxation [25] in the ratio 1:1 was used for updating.

The data for the internal energy $\Delta E$ are given in Table 1. The data for increasing and decreasing $\beta$ are statistically compatible, in other words no "hysteresis" effects are observed. Therefore, we will present graphs for the $1/\sigma^2$-weighted averages only.

We note that $a\Delta E/\beta$ contains a constant contribution

$$c = a\Delta E/\beta|_{\beta=0}\,, \tag{58}$$

leading to a term $c\beta$ in $aM_{\text{in}}$. For the $8^4$ lattice $c = 0.0380$. This contribution is purely magnetic and comes entirely from the part of the fixed plaquettes in the boundary that is attributed to the energy inside the box, cf. the discussion after Eq. (30). In the limit $\beta = 0$ the links become uncorrelated so that the interior does not contribute to $\Delta E/\beta$. For the presentation of the data for $\Delta E$ and $M_{\text{in}}$, however, it will be convenient to ignore the contribution (58). Therefore we will suppress it until we come to the calculation of the physical monopole mass and the function $C$ in Section 4.3.

In Fig. 9 $\Delta E$ is displayed as a function of $\beta$. For $\beta$-values up to 1.8, $\Delta E$ differs little from zero. In this region the correlation length is very small and the boundary conditions do not cause an observable effect in the internal energy, within errors.

The steep increase at $\beta \approx 2.1$ signals that the boundary conditions manage to induce a monopole in the box. As a consequence of the increasing correlation length the interior of the box is then no longer decorrelated from the boundary. It is possible that the appearance of the monopole coincides with the box transition, the analogue of the finite temperature transition for a symmetric hypercube, which is at $\beta_c \approx 2.33$, for a $7^4$ lattice with periodic boundary conditions [26]. The shift from $\beta = 2.33$ to $\approx 2.1$ could be due to the difference in boundary conditions. However, it is also possible that the box/finite temperature transition in $Z_{\text{mon}}$ and $Z_{\text{vac}}$ is not observable in $\Delta E$. This can be studied in large boxes $N \gg N_4$, where one would expect the appearance of the monopole at a $\beta_{\text{mon}}$ much larger than $\beta_c$.

For larger $\beta$, the internal energy slowly decreases towards its asymptotic value, determined by cooling 20 random configurations satisfying the monopole boundary conditions, cf. Section 3.2. This asymptotic value is slightly smaller than the value following from Fig. 7$a$, for $a\mu_0 = \pi/8 \approx 0.39$ ($\mu_0' T/2\pi = 0.75$ in the figure), because the contribution of the plaquettes lying in the boundary is omitted. The physically most interesting region extends from $\beta = 2.3$ onwards, where one expects scaling behaviour.

| $\beta$ | increasing $\beta$ | | | decreasing $\beta$ | | |
|---|---|---|---|---|---|---|
| | $\Delta E_{\mathrm{mag}}$ | $\Delta E_{\mathrm{el}}$ | $\Delta E_{\mathrm{tot}}$ | $\Delta E_{\mathrm{mag}}$ | $\Delta E_{\mathrm{el}}$ | $\Delta E_{\mathrm{tot}}$ |
| 0 | 0 | 0 | 0 | 0 | 0 | 0 |
| 0.20 | $0.041 \pm 0.047$ | $0.028 \pm 0.038$ | $0.069 \pm 0.065$ | $0.057 \pm 0.041$ | $-0.011 \pm 0.044$ | $0.045 \pm 0.061$ |
| 0.40 | $0.017 \pm 0.038$ | $-0.044 \pm 0.030$ | $-0.026 \pm 0.044$ | $-0.028 \pm 0.035$ | $0.011 \pm 0.032$ | $-0.017 \pm 0.043$ |
| 0.60 | $-0.038 \pm 0.033$ | $0.017 \pm 0.034$ | $-0.021 \pm 0.045$ | $-0.015 \pm 0.033$ | $0.032 \pm 0.040$ | $0.017 \pm 0.046$ |
| 0.80 | $0.008 \pm 0.042$ | $0.016 \pm 0.033$ | $0.025 \pm 0.050$ | $-0.058 \pm 0.033$ | $0.003 \pm 0.032$ | $-0.056 \pm 0.045$ |
| 1.00 | $0.052 \pm 0.048$ | $0.018 \pm 0.033$ | $0.070 \pm 0.075$ | $-0.032 \pm 0.035$ | $-0.002 \pm 0.036$ | $-0.034 \pm 0.050$ |
| 1.10 | $0.025 \pm 0.035$ | $0.040 \pm 0.033$ | $0.065 \pm 0.046$ | $0.030 \pm 0.040$ | $0.021 \pm 0.034$ | $0.051 \pm 0.057$ |
| 1.20 | $0.108 \pm 0.040$ | $0.055 \pm 0.032$ | $0.163 \pm 0.050$ | $0.013 \pm 0.035$ | $-0.004 \pm 0.035$ | $0.010 \pm 0.050$ |
| 1.30 | $-0.024 \pm 0.049$ | $-0.018 \pm 0.033$ | $-0.041 \pm 0.061$ | $0.047 \pm 0.039$ | $0.025 \pm 0.040$ | $0.072 \pm 0.064$ |
| 1.40 | $-0.009 \pm 0.042$ | $0.002 \pm 0.040$ | $-0.007 \pm 0.062$ | $-0.001 \pm 0.037$ | $0.068 \pm 0.043$ | $0.067 \pm 0.070$ |
| 1.50 | $0.091 \pm 0.044$ | $0.042 \pm 0.035$ | $0.133 \pm 0.067$ | $-0.013 \pm 0.040$ | $-0.030 \pm 0.043$ | $-0.044 \pm 0.070$ |
| 1.60 | $-0.018 \pm 0.040$ | $-0.022 \pm 0.041$ | $-0.040 \pm 0.065$ | $0.017 \pm 0.037$ | $0.062 \pm 0.037$ | $0.079 \pm 0.063$ |
| 1.70 | $0.036 \pm 0.050$ | $0.016 \pm 0.042$ | $0.052 \pm 0.088$ | $0.020 \pm 0.052$ | $0.045 \pm 0.045$ | $0.065 \pm 0.095$ |
| 1.80 | $0.044 \pm 0.050$ | $0.041 \pm 0.042$ | $0.085 \pm 0.081$ | $-0.025 \pm 0.055$ | $0.017 \pm 0.039$ | $-0.008 \pm 0.084$ |
| 1.90 | $0.128 \pm 0.044$ | $0.083 \pm 0.050$ | $0.211 \pm 0.082$ | $0.100 \pm 0.040$ | $0.056 \pm 0.038$ | $0.156 \pm 0.070$ |
| 1.95* | $0.104 \pm 0.048$ | $0.048 \pm 0.040$ | $0.152 \pm 0.081$ | $0.136 \pm 0.038$ | $0.070 \pm 0.038$ | $0.206 \pm 0.070$ |
| 2.00 | $0.158 \pm 0.050$ | $0.098 \pm 0.050$ | $0.257 \pm 0.097$ | $0.117 \pm 0.044$ | $0.071 \pm 0.040$ | $0.188 \pm 0.080$ |
| 2.05* | $0.159 \pm 0.051$ | $0.123 \pm 0.041$ | $0.282 \pm 0.088$ | $0.121 \pm 0.040$ | $0.071 \pm 0.038$ | $0.191 \pm 0.072$ |
| 2.10* | $0.252 \pm 0.036$ | $0.150 \pm 0.034$ | $0.403 \pm 0.062$ | $0.355 \pm 0.044$ | $0.230 \pm 0.045$ | $0.585 \pm 0.085$ |
| 2.15* | $0.430 \pm 0.039$ | $0.292 \pm 0.032$ | $0.722 \pm 0.067$ | $0.394 \pm 0.055$ | $0.288 \pm 0.045$ | $0.683 \pm 0.090$ |
| 2.20 | $0.457 \pm 0.030$ | $0.334 \pm 0.031$ | $0.791 \pm 0.052$ | $0.468 \pm 0.030$ | $0.319 \pm 0.030$ | $0.786 \pm 0.053$ |
| 2.25* | $0.494 \pm 0.030$ | $0.357 \pm 0.026$ | $0.851 \pm 0.047$ | $0.428 \pm 0.030$ | $0.290 \pm 0.029$ | $0.718 \pm 0.051$ |
| 2.30* | $0.406 \pm 0.026$ | $0.275 \pm 0.026$ | $0.681 \pm 0.043$ | $0.435 \pm 0.030$ | $0.282 \pm 0.027$ | $0.717 \pm 0.049$ |
| 2.35* | $0.461 \pm 0.026$ | $0.309 \pm 0.024$ | $0.770 \pm 0.040$ | $0.461 \pm 0.022$ | $0.305 \pm 0.027$ | $0.766 \pm 0.040$ |
| 2.40 | $0.406 \pm 0.030$ | $0.248 \pm 0.025$ | $0.653 \pm 0.045$ | $0.410 \pm 0.023$ | $0.239 \pm 0.027$ | $0.650 \pm 0.042$ |
| 2.50* | $0.417 \pm 0.018$ | $0.278 \pm 0.021$ | $0.695 \pm 0.030$ | $0.359 \pm 0.019$ | $0.227 \pm 0.021$ | $0.586 \pm 0.030$ |
| 2.60 | $0.371 \pm 0.018$ | $0.218 \pm 0.020$ | $0.589 \pm 0.030$ | $0.393 \pm 0.023$ | $0.190 \pm 0.020$ | $0.583 \pm 0.037$ |
| 2.70* | $0.366 \pm 0.016$ | $0.199 \pm 0.021$ | $0.564 \pm 0.030$ | $0.385 \pm 0.022$ | $0.185 \pm 0.020$ | $0.570 \pm 0.032$ |
| 2.80 | $0.378 \pm 0.015$ | $0.199 \pm 0.016$ | $0.577 \pm 0.024$ | $0.354 \pm 0.015$ | $0.176 \pm 0.016$ | $0.530 \pm 0.023$ |
| 2.90* | $0.367 \pm 0.014$ | $0.176 \pm 0.016$ | $0.542 \pm 0.022$ | $0.381 \pm 0.015$ | $0.185 \pm 0.018$ | $0.566 \pm 0.025$ |
| 3.00 | $0.367 \pm 0.016$ | $0.175 \pm 0.016$ | $0.542 \pm 0.024$ | $0.373 \pm 0.015$ | $0.186 \pm 0.015$ | $0.559 \pm 0.023$ |
| 3.20* | $0.364 \pm 0.013$ | $0.153 \pm 0.015$ | $0.517 \pm 0.021$ | $0.360 \pm 0.012$ | $0.167 \pm 0.013$ | $0.527 \pm 0.019$ |
| 3.50 | $0.364 \pm 0.011$ | $0.151 \pm 0.013$ | $0.516 \pm 0.018$ | $0.361 \pm 0.010$ | $0.141 \pm 0.012$ | $0.503 \pm 0.017$ |
| 4.00 | $0.368 \pm 0.011$ | $0.150 \pm 0.011$ | $0.518 \pm 0.014$ | $0.352 \pm 0.010$ | $0.156 \pm 0.010$ | $0.507 \pm 0.015$ |
| $\infty$ | 0.3841 | 0.1116 | 0.4957 | | | |

Table 1: Monte Carlo data for $\Delta E$, in units of $4\pi/ag^2$, from two cycles of measurements. The magnetic and electric contributions separately as well as the total are given. In the first cycle $\beta$ was gradually increased from 0.2 to 4.0 and subsequently decreased back to 0.2. The second cycle, indicated by the $*$ on the $\beta$ values in the Table, ran from $\beta = 1.95$ to 3.2 and back. Each $\Delta E$ value in the table is based on 4000 configurations. The zero entries for $\beta = 0$ reflect the fact that the contribution $\Delta E/(4\pi/ag^2)|_{\beta=0} = c/\pi = 0.0121$ of Eq. (58) is not included in this table. The $\beta = \infty$ entry is the classical value determined by minimization.

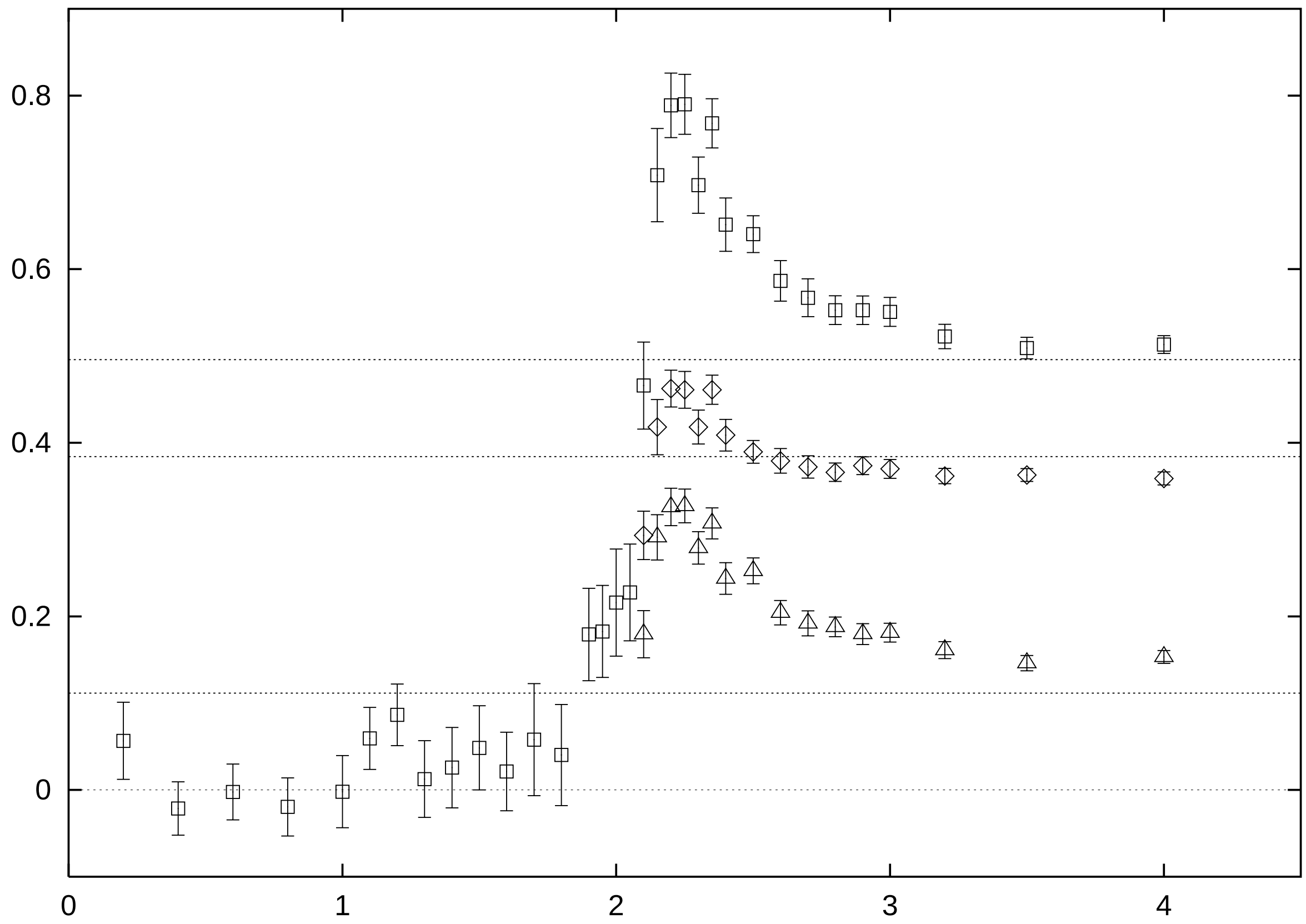

Figure 9: The internal energy $\Delta E$ in units of $4\pi/ag^2$ as a function of $\beta = 4/g^2$. Shown are the magnetic ($\diamond$) and electric ($\triangle$) components and the total ($\square$). For $\beta < 2.1$ only the total is plotted for clarity. The dotted horizontal lines indicate the asymptotic ($\beta \to \infty$) values, determined by minimization, cf. Table 1.

The magnetic and electric components of the internal energy are seen to follow essentially the same behaviour as a function of $\beta$. The fact that the asymptotic, classical, value of the electric energy is much smaller than that of the magnetic energy is due to the small value of $\mu_0 R$ (cf. Fig. 7$a$ at $\mu_0' T/2\pi = 0.75$, which corresponds to the same value of $\mu_0 R$). The magnetic and electric internal energies do not differ as much in the intermediate $\beta$ region as in the limit $\beta \to \infty$. As a matter of fact, the magnetic energy approaches its limiting value very slowly, and from below, whereas the electric contribution approaches its asymptotic value from above.

It is also interesting to look at the spatial density profile of the internal energy. It is shown in Fig. 10 for several values of $\beta$. For clarity we present it without error bars and with data points connected by straight lines to guide the eye. The energy density is given in units of the inverse lattice spacing $a$. However, since $a\mu_0$ is fixed at $\pi/8$, and $R/a = 3.5$, this is equivalent to units of either of the monopole scales $\mu = \mu_0$ or $\mu = \mu_0 + 1/R_{\rm eff}$. The energy density at $r$ is computed by averaging the energies of all elementary cubes whose centre is at a distance $r$ from the centre of the box. The energy in such a cube is proportional to the sum of the plaquettes forming it. The resulting $r$- dependent distribution is averaged over time slices and Monte Carlo configurations. The figure shows clearly that a lump of energy builds up when the monopole is induced in the box for $\beta \gtrsim 2.2$. With increasing $\beta$, the energy density decreases towards the classical $\beta \to \infty$ curve, in accordance with the observation made for the total internal energy, Fig. 9. The wiggles in the classical curve give an indication of the size of discretization effects and of boundary effects due to the cubic geometry.

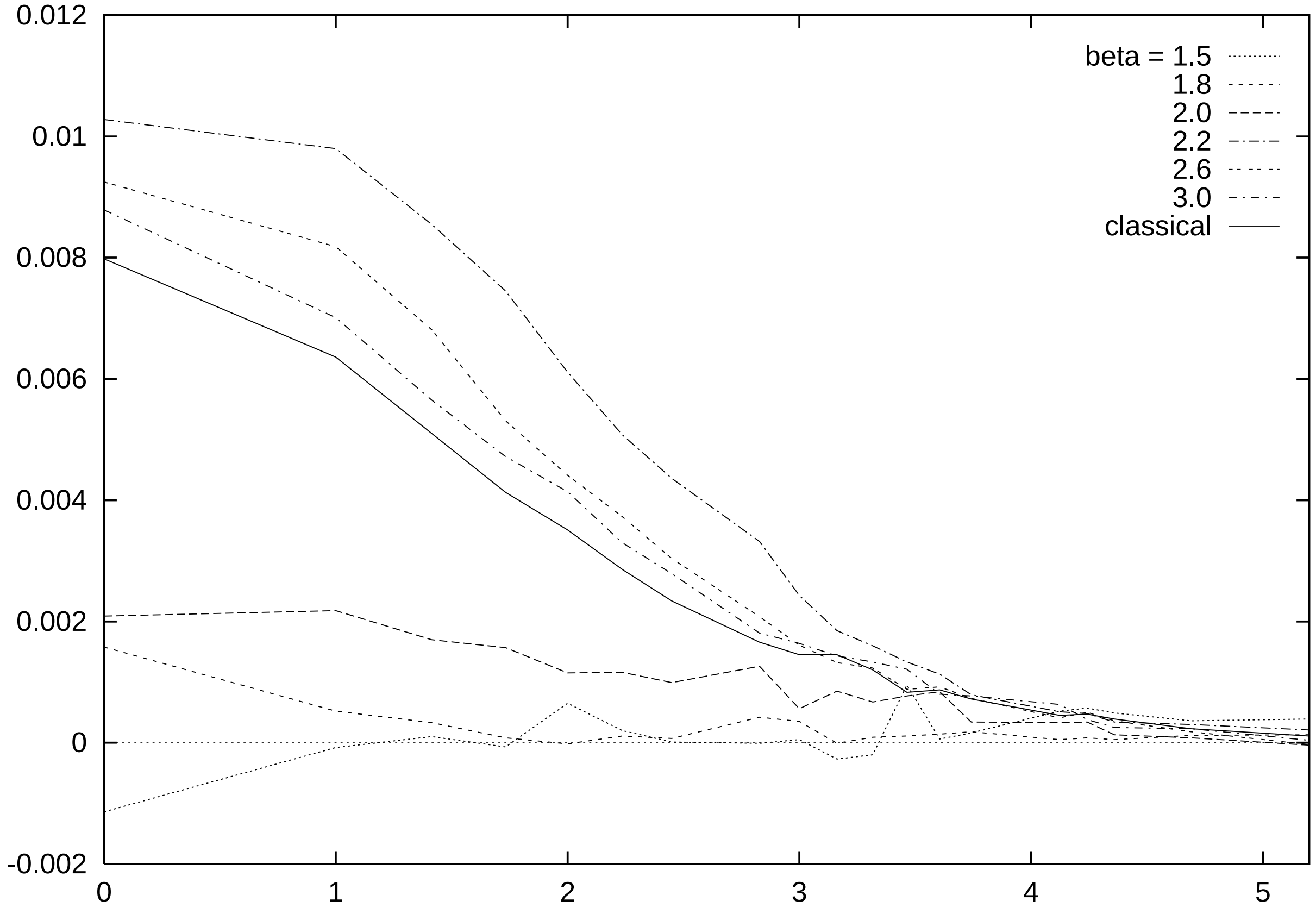

Figure 10: Expectation values for the internal energy density in units of $4\pi/a^4g^2$, as a function of $r/a$.

The behaviour of the energy density as a function of $\beta$ does not provide conclusive information as to whether the quantum monopole becomes HP-like or stays BPS like. (The data for electric and magnetic contributions separately are also inconclusive.) Note, for comparison, that the finite temperature electric screening mass $\sqrt{2/3}\, g_R/aN_4$ is fairly small for couplings $g_R^2 \lesssim 10$.

### 4.3 The monopole mass and the function $C(g_R^2)$

Now we turn to the computation of the monopole mass. The results for $M_{\rm in}$ in units of the lattice spacing are shown in Fig. 11. These data have been obtained by integrating the data for the internal energy $\Delta E$ of Table 1 and Fig. 9, using Eq. (57). The constant contribution $c\beta$ from the boundary plaquettes (58) is still omitted. At each given $\beta$-value, the integration was carried out for all the 4000 configurations separately, allowing an estimate of the statistical errors on the computed masses by binning. Because of the integration procedure, the masses at neighbouring values of $\beta$ are strongly correlated.

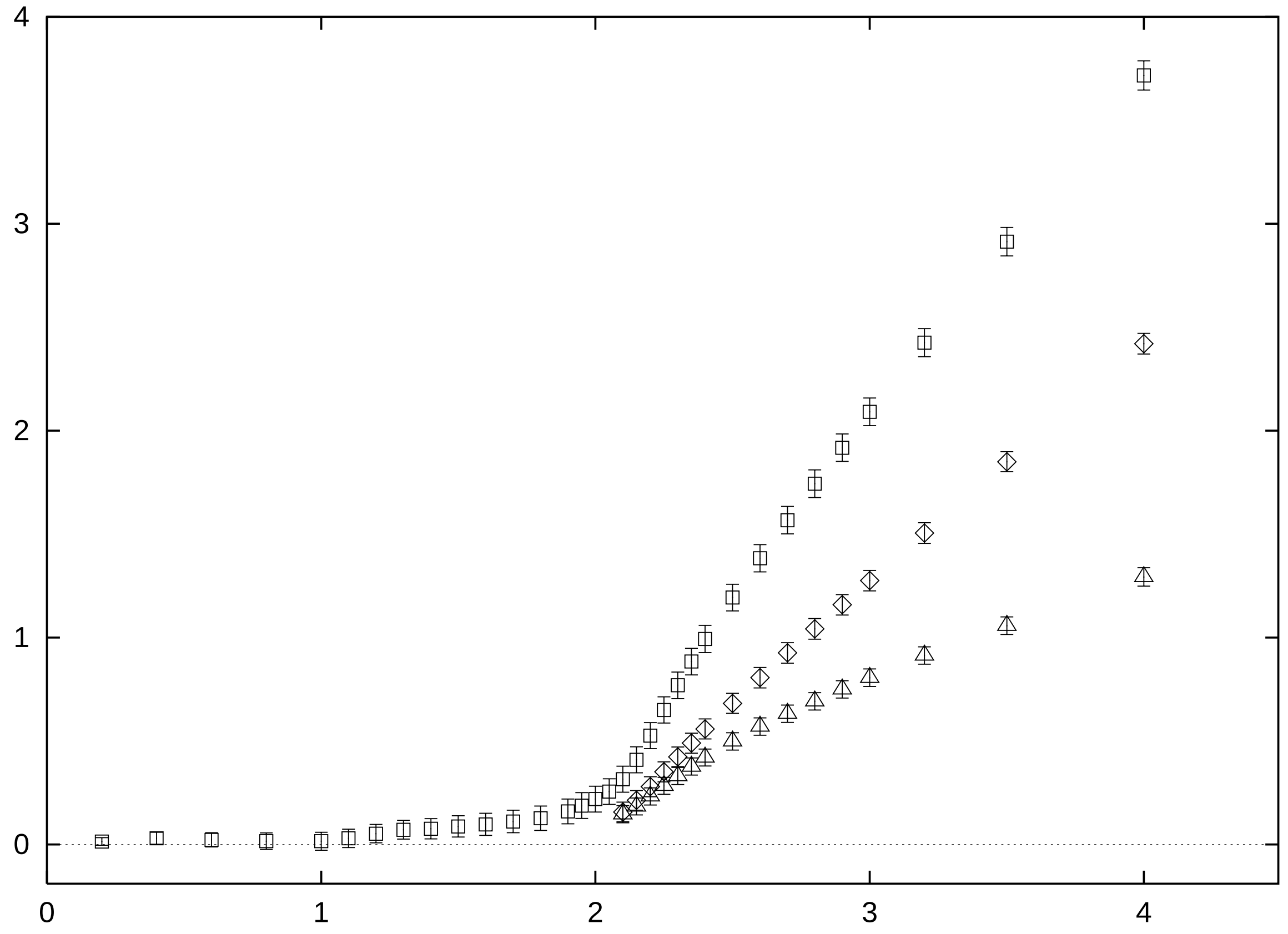

Figure 11: The dimensionless monopole mass $aM_{\rm in}$ as a function of $\beta$, for $a\mu_0 = \pi/8$. Apart from the total mass ($\square$), the magnetic ($\diamond$) and electric ($\triangle$) contributions are shown separately. For $\beta < 2.1$ only the total mass is plotted for clarity.

In the limit $\beta \to \infty$, the mass curve approaches a straight line with slope given by $a\Delta E(\beta \to \infty) = 0.4957 \times 4\pi/g^2 = 1.5573\,\beta$, see Fig. 9. This line is approached from below because $\Delta E$ is a decreasing function at large $\beta$. A fit of the mass data in the large-$\beta$ region, including the $c\beta$ contribution, gives

$$aM_{\rm in}(\beta) \approx 1.5953\,(\beta - 1.415) - 0.97\,\frac{1}{\beta}\,, \tag{59}$$

with prescribed slope $1.5573 + 0.0380 = 1.5953$.

We now use the mass data to compute the function $C$ in the mass formula Eq. (2). For the interior of the box we write

$$aM_{\rm in}(\beta; a\mu_0) = \frac{4\pi\, a\mu}{g_R^2(\Lambda_R/\mu)}\, C_{\rm in}(g_R^2(\Lambda_R/\mu))\,, \tag{60}$$

where we do the analysis for the HP and BPS values of $\mu$ given in (48,49). In order to extract $C_{\rm in}$ from the mass data, using this formula, we need to compute $\Lambda_R/\mu$ for both choices of $\mu$ and subsequently $g_R^2(\Lambda_R/\mu)$ as a function of $\beta$. Then the total $C$ is obtained by taking into account the correction (52) or (51) for the outside region.

We use lattice string tension data to determine the lattice distance $a$ in terms of a physical scale. We write

$$\frac{\mu}{\Lambda_R} = a\mu \, \frac{1}{a\sqrt{\sigma}} \, \frac{\sqrt{\sigma}}{\Lambda_R} \, . \tag{61}$$

The quantity $\sqrt{\sigma}/\Lambda_R$ is an input parameter, since the Monte Carlo simulations have not yet determined it unambiguously. The data on the string tension $a\sqrt{\sigma}$ were collected from various sources [27–29]. We interpolated some of them towards neighbouring $\beta$-values. The values we used, tabulated in Table 2, come from Polyakov-loop correlations and include the finite-size correction $\pi/3L^2$ to $\sigma$ [30–32]. Possible systematic errors on $\sigma$ are inherited by our function $C$. We have analysed our data for three different values of $\sqrt{\sigma}/\Lambda_R$: 1.7, 2.0 and 2.2. The former is consistent with both the value given in Ref. [33] and the value obtained by combining the $\beta = 2.85$ result $a\sqrt{\sigma} = 0.0636$ of Ref. [29] with the lower bound 0.038 on $a\Lambda_R$ given there. The value 2.0 for $\sqrt{\sigma}/\Lambda_R$ follows from Ref. [34], and 2.2 is the variational estimate we obtained in Ref. [1].

The running coupling $g_R^2(\Lambda_R/\mu)$ is determined from the $\Lambda_R/\mu$ values thus obtained by means of the two-loop $\beta$-function for this coupling,

$$-\frac{1}{g_R}\beta(g_R) = -\frac{\mu}{g_R}\frac{\partial g_R}{\partial \mu} = b_0 g_R^2 + b_1 g_R^4 \, , \tag{62}$$

with coefficients $b_0 = 11/(24\pi^2)$, $b_1 = (102/121)\, b_0^2$ for SU(2). The exact solution at this order in $g_R^2$ is given by

$$\frac{\mu}{\Lambda_R} = \left( \frac{1}{b_0 g_R^2} + \frac{b_1}{b_0^2} \right)^{-b_1/2b_0^2} \exp\left( \frac{1}{2b_0 g_R^2} \right) \, . \tag{63}$$

We find $g_R^2(\Lambda_R/\mu)$ by inverting Eq. (63) numerically, instead of resorting to an approximate analytic inversion. The use of the two-loop formula becomes questionable at distance scales of the order $1/\sqrt{\sigma}$, at the 'border of the semiclassical region', when $g_R^2$ becomes large. We keep in mind, therefore, that this may cause a systematic error of perhaps 20% in our HP-results at $\beta = 2.3$.

This analysis leads to the values for $C$ given in Table 2, where we have included the contribution $c\beta = 0.0380\,\beta$ from the plaquettes in the boundary (58). The quoted numbers take into account $C_{\rm out}^{\rm HP} = 0.302$ or $C_{\rm out}^{\rm BPS} = 0.368$, from (52,51) and (48,49). The $\beta = \infty$ entries were obtained from $C_{\rm in} = \Delta E/(\pi\beta\mu)$, $\beta \to \infty$, giving $C_{\rm in}^{\rm HP} = (0.4957\pi + 0.0380)/(0.393\,\pi) = 1.293$, $C^{\rm HP} = 1.596$, and $C_{\rm in}^{\rm BPS} = (0.393/0.646)\, C_{\rm in}^{\rm HP} = 0.787$, $C^{\rm BPS} = 1.155$. These $C$-values are useful for reference, although at $\beta = \infty$ the classical pure SU(2) monopole is BPS with $C = 1$.

Fig. 12 shows $C$ as a function of $g_R^2(\Lambda_R/\mu)$. Since $\mu_{\rm BPS} > \mu_{\rm HP}$ the points in the BPS analysis are found at weaker running coupling than those in the HP analysis. The dotted horizontal lines are the $\beta \to \infty$ values 1.596 and 1.155 described above. The error bars correspond to the statistical errors on the mass data only. The data points in each set should lie on a smooth curve because they are strongly correlated, as in the case of the mass data of Fig. 11, and the error bars represent the interval within which these curves as a whole can move. The small deviations of the data points themselves from the smooth behaviour of the dashed lines in Fig. 12, drawn to guide the eye, are due to errors in the string tension data used.

| $\beta$ | $a\sqrt{\sigma}$ | $aM_{\rm in}$ | $\sqrt{\sigma}/\Lambda_R = 1.7$ | | $\sqrt{\sigma}/\Lambda_R = 2.0$ | | $\sqrt{\sigma}/\Lambda_R = 2.2$ | |
|---|---|---|---|---|---|---|---|---|
| | | | $g_R^2$ | $C$ | $g_R^2$ | $C$ | $g_R^2$ | $C$ |
| 2.20 | 0.467 | $0.609 \pm 0.063$ | 15.46 | $1.95 \pm 0.20$ | 11.42 | $1.52 \pm 0.15$ | 9.99 | $1.37 \pm 0.13$ |
| 2.25 | 0.415 | $0.735 \pm 0.063$ | 12.26 | $1.92 \pm 0.16$ | 9.70 | $1.58 \pm 0.12$ | 8.69 | $1.45 \pm 0.11$ |
| 2.30 | 0.368 | $0.856 \pm 0.064$ | 10.25 | $1.90 \pm 0.13$ | 8.46 | $1.62 \pm 0.11$ | 7.70 | $1.50 \pm 0.10$ |
| 2.35 | 0.313 | $0.973 \pm 0.064$ | 8.47 | $1.82 \pm 0.11$ | 7.26 | $1.60 \pm 0.09$ | 6.71 | $1.50 \pm 0.09$ |
| 2.40 | 0.266 | $1.084 \pm 0.066$ | 7.25 | $1.76 \pm 0.10$ | 6.37 | $1.58 \pm 0.08$ | 5.95 | $1.50 \pm 0.08$ |
| 2.50 | 0.186 | $1.288 \pm 0.064$ | 5.58 | $1.65 \pm 0.07$ | 5.06 | $1.53 \pm 0.07$ | 4.81 | $1.46 \pm 0.06$ |
| 2.60 | 0.132 | $1.482 \pm 0.066$ | 4.60 | $1.59 \pm 0.06$ | 4.25 | $1.49 \pm 0.06$ | 4.07 | $1.44 \pm 0.05$ |
| 2.70 | 0.101 | $1.670 \pm 0.067$ | 4.06 | $1.59 \pm 0.05$ | 3.79 | $1.51 \pm 0.05$ | 3.65 | $1.46 \pm 0.05$ |
| 2.80 | 0.0742 | $1.850 \pm 0.067$ | 3.58 | $1.57 \pm 0.05$ | 3.37 | $1.49 \pm 0.05$ | 3.26 | $1.45 \pm 0.04$ |
| 2.90 | 0.0545 | $2.027 \pm 0.067$ | 3.21 | $1.55 \pm 0.04$ | 3.04 | $1.48 \pm 0.04$ | 2.95 | $1.45 \pm 0.04$ |
| 3.00 | 0.0400 | $2.205 \pm 0.067$ | 2.91 | $1.53 \pm 0.04$ | 2.77 | $1.48 \pm 0.04$ | 2.70 | $1.45 \pm 0.04$ |
| $\infty$ | 0 | $\infty$ | 0 | 1.596 | 0 | 1.596 | 0 | 1.596 |

| $\beta$ | $a\sqrt{\sigma}$ | $aM_{\rm in}$ | $\sqrt{\sigma}/\Lambda_R = 1.7$ | | $\sqrt{\sigma}/\Lambda_R = 2.0$ | | $\sqrt{\sigma}/\Lambda_R = 2.2$ | |
|---|---|---|---|---|---|---|---|---|
| | | | $g_R^2$ | $C$ | $g_R^2$ | $C$ | $g_R^2$ | $C$ |
| 2.20 | 0.467 | $0.609 \pm 0.063$ | 7.70 | $0.95 \pm 0.06$ | 6.70 | $0.87 \pm 0.05$ | 6.24 | $0.84 \pm 0.05$ |
| 2.25 | 0.415 | $0.735 \pm 0.063$ | 6.94 | $1.00 \pm 0.05$ | 6.14 | $0.92 \pm 0.05$ | 5.75 | $0.89 \pm 0.04$ |
| 2.30 | 0.368 | $0.856 \pm 0.064$ | 6.33 | $1.04 \pm 0.05$ | 5.66 | $0.97 \pm 0.04$ | 5.33 | $0.93 \pm 0.04$ |
| 2.35 | 0.313 | $0.973 \pm 0.064$ | 5.66 | $1.05 \pm 0.04$ | 5.13 | $0.98 \pm 0.04$ | 4.87 | $0.95 \pm 0.04$ |
| 2.40 | 0.266 | $1.084 \pm 0.066$ | 5.13 | $1.05 \pm 0.04$ | 4.70 | $1.00 \pm 0.04$ | 4.48 | $0.97 \pm 0.04$ |
| 2.50 | 0.186 | $1.288 \pm 0.064$ | 4.27 | $1.05 \pm 0.03$ | 3.97 | $1.00 \pm 0.03$ | 3.82 | $0.97 \pm 0.03$ |
| 2.60 | 0.132 | $1.482 \pm 0.066$ | 3.69 | $1.04 \pm 0.03$ | 3.47 | $1.00 \pm 0.03$ | 3.35 | $0.98 \pm 0.03$ |
| 2.70 | 0.101 | $1.670 \pm 0.067$ | 3.34 | $1.06 \pm 0.03$ | 3.16 | $1.02 \pm 0.03$ | 3.06 | $1.00 \pm 0.03$ |
| 2.80 | 0.0742 | $1.850 \pm 0.067$ | 3.02 | $1.06 \pm 0.02$ | 2.87 | $1.02 \pm 0.02$ | 2.79 | $1.00 \pm 0.02$ |
| 2.90 | 0.0545 | $2.027 \pm 0.067$ | 2.75 | $1.06 \pm 0.02$ | 2.63 | $1.03 \pm 0.02$ | 2.57 | $1.01 \pm 0.02$ |
| 3.00 | 0.0400 | $2.205 \pm 0.067$ | 2.53 | $1.06 \pm 0.02$ | 2.43 | $1.03 \pm 0.02$ | 2.37 | $1.01 \pm 0.02$ |
| $\infty$ | 0 | $\infty$ | 0 | 1.155 | 0 | 1.155 | 0 | 1.155 |

Table 2: Computed values for $g_R^2$ and $C$, for the HP analysis (top) and the BPS analysis (bottom). The analyses have been done for the three quoted values of $\sqrt{\sigma}/\Lambda_R$, as described in the text. The values for the string tension used for the calculation, obtained from Refs. [27–29], are given as well.

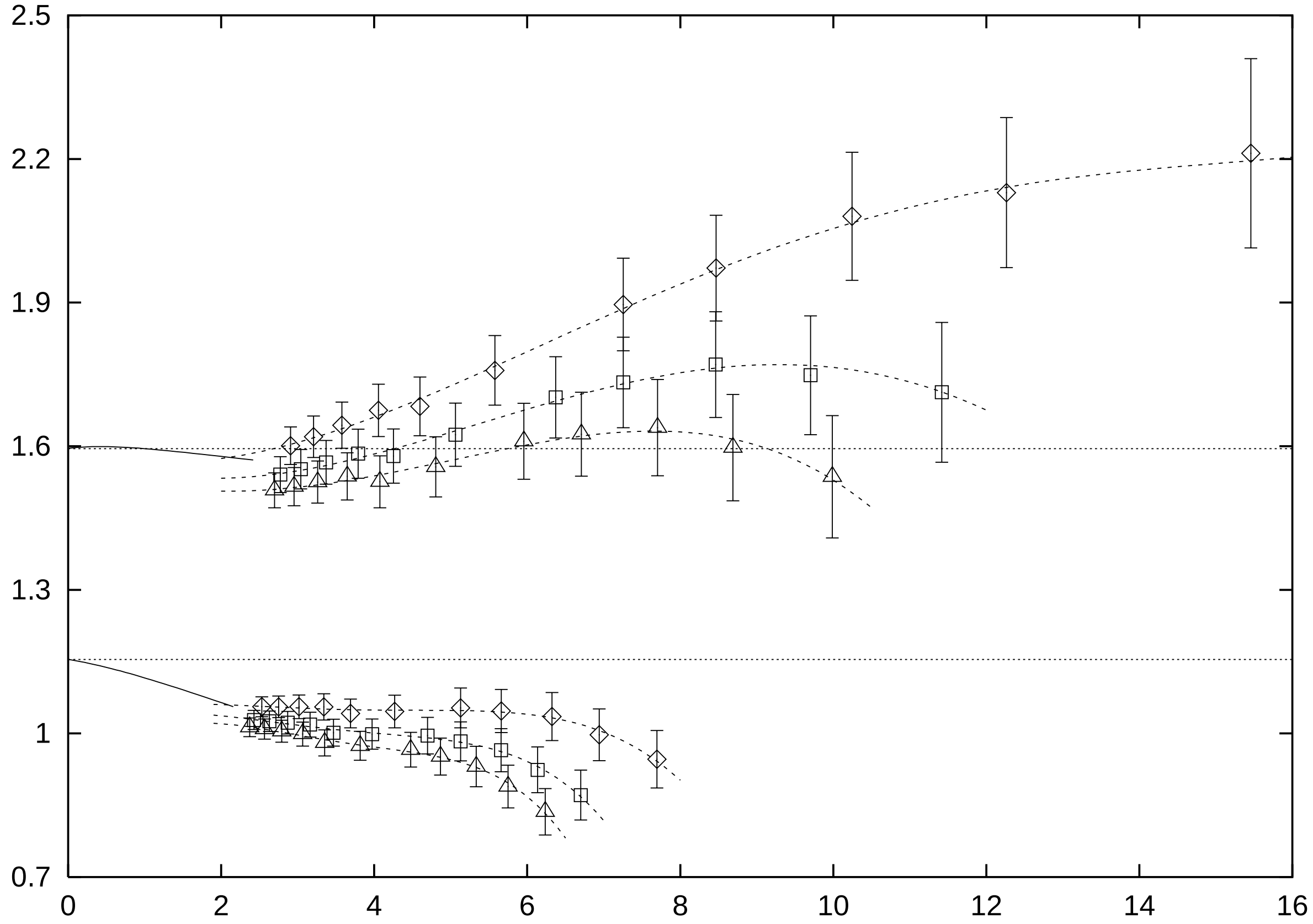

Figure 12: $C$ as a function of $g_R^2(\Lambda_R/\mu)$. The different sets of points are for $\sqrt{\sigma}/\Lambda_R = 1.7$ ($\diamond$), 2.0 ($\square$) and 2.2 ($\triangle$), for both the HP analysis (upper set of curves) and the BPS analysis (lower set). The eleven points for each value of $\sqrt{\sigma}/\Lambda_R$ and $\mu$ are based on the mass data for $\beta = 2.2$, 2.25, 2.3, 2.35, 2.4, 2.5, 2.6, 2.7, 2.8, 2.9, 3.0, from right to left.

The solid curves in Fig. 12 give a rough indication of the approach to the classical limit. They have been obtained from the fit (59). Since there are no string tension data available for large $\beta$, we used the 2-loop approximation to the $\beta$-function for the bare lattice coupling here to compute $a\Lambda_L$ and hence $\Lambda_R/\mu$ for both values of $\mu$, using $\Lambda_R/\Lambda_L = 20.78$ [10]. It is well known that this is a bad approximation for intermediate $\beta$, so these curves should be viewed with caution.

In the small-$g_R^2$ region the monopole will be BPS-like. The deviation of the classical $C$-value of 1.155 from the true value $C = 1$ of the classical BPS monopole is primarily an 'exponential correction' due to the small value or $\mu R$ in our simulation. To a lesser extent, it is also caused by our use of HP boundary conditions for simulating the BPS monopole (cf. the slight difference between HP and BPS in Fig. 8). Thus, the small-$g_R^2$ points in the HP analysis in Fig. 12 are misleading. At larger values of $g_R^2$ we have two scenarios. If the monopole remains BPS-like, one of the lower curves applies, depending on the value of $\sqrt{\sigma}/\Lambda_R$. If the monopole becomes HP-like, on the other hand, the graph of $C$ will lie in between the two sets of curves in the plot, starting off at $C = 1$ and rising to one of the upper curves at intermediate or large $g_R^2$.

In the HP scenario, the conclusion we draw is that the mass formula (2) holds over a wide range of $g_R^2$ values. Using only values of $\beta$ in the scaling regime $\beta \geq 2.3$, and taking a BPS monopole in the $\beta \to \infty$ limit as discussed above, we see that the function $C$ is slowly increasing as a function of $g_R^2$. In the BPS scenario, we had no conjecture about the behaviour of $C$. (Recall that we introduced this function because of the analogy with the mass formula for the 't Hooft-Polyakov monopole.) Thus, there is no reason to expect it to be an increasing function of $g_R^2$. Nevertheless, even in this case $C$ varies little, albeit over a smaller range of running couplings.

# 5 Conclusion

We have performed a numerical investigation of the magnetic monopole in pure SU(2) gauge theory. The classical monopoles could be caught as local minima on a lattice with prescribed boundary conditions. Gauge symmetries limited the scales $\mu$ of the monopoles corresponding to global minima, relevant for the Monte Carlo simulations in the quantum theory. The preferred scale was related to the finite extent of the system in the temperature direction. Since the boundary conditions induce the quantum monopole only for finite volume, we were led to studying the interior of the monopole in a finite temperature situation. The exploratory numerical data for the interior of the monopole show how the energy density (Fig. 10) and mass (Fig. 11) behave as a function of the bare gauge coupling.

In view of our earlier work [1] we analyzed these data with two scenarios: the quantum monopole could remain BPS-like or become HP-like. Our data for the energy density were insufficient to decide between the two alternatives. Assuming that the monopole behaves HP-like at intermediate and large values of the coupling, we found the function $C(g_R^2)$ in (2) to be slowly increasing, for renormalized coupling values ranging from $g_R^2 \approx 3$ to 8 ($C \approx 1.6$), or 3 to 10 ($1.6 \lesssim C \lesssim 2.0$), depending on the value of $\sqrt{\sigma}/\Lambda_R$ (cf. Fig. 12). These coupling values correspond to scales $\mu$ varying from about $10\sqrt{\sigma}$ to $\sqrt{\sigma}$, which include the $\mu$ of order $\sqrt{\sigma}$ needed in Ref. [1].

As a variant of the HP interpretation we can take the simulated monopole at face value. It is an object with magnetic charge, and a free energy given by the mass formula. The shape of the energy density suggests that these monopoles keep their identity when packing them at a minimal distance $b = \mu^{-1}$.

Even in the BPS scenario the mass formula appears to have reasonable validity. The corresponding function $C$ was found to decrease only slightly between $g_R^2 = 0$ and 5.5 or 7.

Hence, in both scenarios the scale dependence of the monopole mass is essentially given by the scale dependence of the renormalized coupling. This supports our assumption about the monopole mass mentioned in the introduction.

These results can of course be improved by going to larger lattices. In particular, controlled matching with the classical limit $C = 1$ for $g_R \to 0$, which was not possible with our small spatial volume, may be possible using asymmetric $N^3 \times N_4$ lattices with $N \gg N_4$. This may also allow a more detailed investigation of the screening properties of the electric charge.

## Acknowledgements

This work was supported by SERC grant GR/H01243, by the Stichting voor Fundamenteel Onderzoek der Materie (FOM), and by DGICYT. The numerical simulations were performed on the Cray Y-MP4/464 at SARA and were supported by the Stichting Nationale Computer Faciliteiten (NCF).